\documentclass[12pt]{article}
\newcommand{\be}{\begin{equation}}
\newcommand{\ee}{\end{equation}} 

\newcommand{\bi}{\begin{itemize}}
\newcommand{\ei}{\end{itemize}}
\newcommand{\bea}{\begin{eqnarray}}
\newcommand{\eea}{\end{eqnarray}}
\newcommand{\ba}{\begin{array}}
\newcommand{\ea}{\end{array}}

\usepackage[left=2.50cm, right=2.50cm, top=2.50cm, bottom=2.50cm]{geometry}
\usepackage[utf8]{inputenc}
\usepackage{cancel}
\usepackage[english]{babel}
\usepackage{physics}
\usepackage{hyperref}
\usepackage{amsthm}
\usepackage{mathrsfs}
\usepackage{mathtools}
\usepackage{graphicx}
\usepackage{changepage}
\usepackage{amssymb,amsmath}
\usepackage{verbatim}
\usepackage{empheq}
\usepackage{xcolor}
\usepackage{tikz}
\usepackage{float}
\usepackage{multirow}
\usepackage{multicol}
\usepackage{amsmath}
\usepackage{tensor}
\usepackage{centernot}
\usepackage[normalem]{ulem}
\usepackage[hang,flushmargin]{footmisc} 
\usepackage[normalem]{ulem}

\newcommand{\invmixt}[3]{\tensor{#1}{_{#2}^{#3}}}
\usetikzlibrary{tikzmark}
\numberwithin{equation}{section}
\hypersetup{hidelinks}

\newlength{\bibitemsep}
\newlength{\bibparskip}
\let\oldthebibliography\thebibliography
\renewcommand\thebibliography[1]{%
  \oldthebibliography{#1}%
  \setlength{\parskip}{\bibitemsep}%
  \setlength{\itemsep}{\bibparskip}%
}
\begin{document}
\par
\bigskip
\Large
\noindent
{\bf Fractons from covariant higher-rank 3D BF theory\\
\par
\rm
\normalsize

\hrule

\vspace{0.7cm}
\large
\noindent
{\bf Erica Bertolini$^{1,a}$},
{\bf Alberto Blasi$^{2,b}$}, 
{\bf Matteo Carrega$^{3,c}$},\\
{\bf Nicola Maggiore$^{2,4,d}$},
{\bf Daniel Sacco Shaikh$^{4,e}$}\\

\par

\small

\noindent$^1$ School of Theoretical Physics, Dublin Institute for Advanced Studies, 10 Burlington Road, Dublin 4, Ireland.

\noindent$^2$ Istituto Nazionale di Fisica Nucleare - Sezione di Genova, Via Dodecaneso 33, I-16146 Genova, Italy.

\noindent$^3$ CNR-SPIN, Via Dodecaneso 33, 16146 Genova, Italy.

\noindent$^4$ Dipartimento di Fisica, Universit\`a di Genova, Via Dodecaneso 33, I-16146 Genova, Italy.
\smallskip

\smallskip

\vspace{1cm}

\noindent
{\tt Abstract}\\
In this paper we study the 3D gauge theory of two tensor gauge fields: $a_{\mu\nu}(x)$, which we take symmetric, and $B_{\mu\nu}(x)$, with no symmetry on its indices. The corresponding invariant action is a higher-rank BF-like model, which is first considered from a purely field theoretical point of view, and the propagators with their poles and the degrees of freedom are studied. Once matter is introduced, a  fracton behaviour naturally emerges. We show that our theory can be mapped to the low-energy effective field theory describing the Rank-2 Toric Code (R2TC). This relation between our covariant BF-like theory and the R2TC is a higher-rank generalization of the equivalence between the ordinary 3D BF theory and the Kitaev's Toric Code. In the last part of the paper we analyze the case in which the field $B_{\mu\nu}(x)$ is a symmetric tensor. It turns out that the obtained BF-like action can be cast into the sum of two rank-2 Chern-Simons actions, thus generalizing the ordinary abelian case. Therefore, this represents a higher-rank generalization of the ordinary 3D BF theory, which well describes the low-energy physics of quantum spin Hall insulators in two spatial dimensions.

\vspace{\fill}

\noindent{\tt Keywords:} \\
Quantum field theory, tensor gauge field theory, BF theory, fractons.

\vspace{1cm}

\hrule
\noindent{\tt E-mail:
$^a$ebertolini@stp.dias.ie,
$^b$alberto.blasi@ge.infn.it,
$^c$matteo.carrega@spin.cnr.it,
$^d$nicola.maggiore@ge.infn.it,
$^e$saccodaniel@outlook.it.
}
\newpage


\section{Introduction}

From the middle of last century, quantum field theory has allowed us to enrich our modern comprehension of strongly correlated systems \cite{Fradkin:2013sab}, characterized by a macroscopic number of interacting Degrees of Freedom (DoF) whose collective behaviour gives rise to emergent phenomena which are not expected by studying the single microscopic components only \cite{Anderson:1972pca}. For instance, many quantum phases of matter \cite{Sachdev:2011fcc}, such as quantum Hall fluids \cite{Zee:1995avy} and quantum spin liquids \cite{Fradkin:2013sab}, can be described in terms of vector gauge theories. As an example, Chern-Simons \cite{Dunne:1998qy} and BF theories \cite{Birmingham:1991ty, Blasi:2011pf} have been shown to be the natural theoretical frameworks to describe, respectively, the quantum Hall effect \cite{Tong:2016kpv} and the topological insulators \cite{Hasan:2010xy, Cho:2010rk}. On the other hand, many discoveries and ideas first developed in condensed matter and statistical physics had a later impact in the investigation of new fundamental problems of quantum field theory \cite{Fradkin:2013sab}. This is exactly the case of fractons, exotic emergent excitations of certain quantum phases of matter which are completely immobile in isolation \cite{Vijay:2015mka, Prem:2017kxc, Pretko:2020cko, Gromov:2022cxa}.\\

 Fractons represent an extreme case of subdimensional particles, \textit{i.e.} particles which are restricted to move only in lower dimensional subspaces. In particular,  particles which can only move on zero-, one- and two-dimensional subspaces are known, respectively, as fractons, lineons and planons \cite{Pretko:2020cko}.
 \textcolor{black}{The first realization of this subdimensional behaviour has been made} in exactly solvable quantum spin models \textcolor{black}{with discrete symmetries} as quantum error-correcting codes \cite{Chamon:2004lew, Haah:2011drr, Vijay:2016phm}. The prototypical examples of fracton spin models in  three spatial dimensions are the X-Cube \cite{Vijay:2016phm, Slagle:2017wrc} and the Haah's Code \cite{Haah:2011drr} which belong, respectively, to ``type I'' and ``type II'' classes, the former being characterized by all three types of subdimensional excitations while the latter has fractons only. 
Since then, fractons have attracted more and more attention in many different areas of theoretical physics, such as quantum field theory \cite{Seiberg:2020bhn, Huang:2023zhp}, elasticity \cite{Pretko:2017kvd, Gromov:2017vir}, hydrodynamics \cite{Gromov:2020yoc, Grosvenor:2021rrt, Doshi}, and gravity \cite{Pretko:2017fbf, Blasi:2022mbl, Afxonidis:2023pdq}.\\

Remarkably, it has been proved that fractons are well-described in the language of  \textcolor{black}{$U(1)$} rank-2 symmetric tensor gauge theories \cite{Pretko:2020cko, Pretko:2016lgv}. For example, in the so-called ``scalar charge  theory'' \cite{Pretko:2016lgv}, the limited mobility property of fractons is achieved
through a Gauss-like law
\begin{equation}
    \partial_ i \partial_j E^{ij} = \rho \,,\label{constraint}
\end{equation}
where $E^{ij}(x)$ is a generalized rank-2 symmetric electric tensor field and $\rho(x)$ is the charge density.
From the constraint \eqref{constraint} it follows immediately that both the total charge and the total dipole momentum are conserved up to boundary contributions, hence isolated single particles cannot move, which is the defining property of fractons.
In the framework of gauge theories, the constraint \eqref{constraint} can be derived from a generalized electromagnetism for a tensor gauge field  $a_{ij}(x)$ transforming according to the following rank-2 generalization of the usual $U(1)$ gauge transformation \cite{Pretko:2016lgv}
\begin{equation}
    \delta a_{ij} = \partial_i \partial_j \Lambda\ ,
\end{equation}
where $\Lambda(x)$ is a local scalar function and ($i,j$) are spatial indices. 
It is interesting that the only mobile low-energy excitations are dipoles (\textit{i.e.} bound states of fractons) whose motion is represented by a rank-2 symmetric tensor current $J^{ij}(x)$, which is related to the fractonic  density $\rho(x)$ through the continuity equation
\begin{equation}
    \partial_0 \rho + \partial_i \partial_j J^{ij}=0\ ,
\end{equation}
encoding the conservation of total charge and dipole momentum. \\

 Recently, a higher-rank generalization of the Kitaev's Toric Code \cite{Kitaev:1997wr} has been investigated, realizing the Rank-2 Toric Code (R2TC) in two spatial dimensions \cite{Oh:2021gee, Oh:2022klh, Pace:2022wgl, Oh:2023bnk}, which is an exactly solvable quantum lattice model whose excitations exhibit restricted mobility and unusual braiding statistics. In particular, the phase factor acquired by the wavefunction in the braiding of an excitation around another one depends on the initial positions of the quasiparticles involved \cite{Oh:2021gee}. Furthermore, such a phase captures the total dipole momentum of the excitations enclosed in the process, differently from what happens in standard 3D Chern-Simons and BF theories, where, instead, the total charges are involved \cite{Oh:2022klh}. The R2TC has a discrete symmetry group $\mathbb{Z}_N$ and it can be obtained from the rank-2 $U(1)$ lattice gauge theory, whose excitations are subdimensional, through the Higgsing procedure \cite{Bulmash:2018lid}, as shown in \cite{Oh:2021gee}, which lowers the $U(1)$ continuous gauge symmetry to $\mathbb{Z}_N$. The low-energy effective field theory embodies dipole symmetry and it has been shown to be equivalent to a dipolar BF description \cite{Ebisu:2023idd}  in the context of foliated theories \cite{Slagle:2020ugk}. \\

From a field theoretical point of view, it has been recently shown (both in 3D and 4D) that covariant theories of fractons \cite{Blasi:2022mbl, Bertolini:2022ijb, Bertolini:2023juh, Bertolini:2023sqa, Bertolini:2024yur} can be obtained by studying the most general power counting compatible action of a rank-2 symmetric tensor field $a_{\mu\nu}(x)$ ($\mu,\nu$ spacetime indices) with mass dimension one, and invariant under the covariant gauge transformation
\begin{equation}
    \delta a_{\mu\nu} = \partial_\mu \partial_\nu \Lambda\,, \label{longitudinal_diffeomorphisms}
\end{equation}
also known as longitudinal diffeomorphisms \cite{Dalmazi:2020xou}. In the 4D case the resulting theory consists of two terms: the first is linearized gravity \cite{Hinterbichler:2011tt,Gambuti:2020onb, Gambuti:2021meo,Bertolini:2023wie} (which is expected since \eqref{longitudinal_diffeomorphisms} is a particular case of the diffeomorphism symmetry) while the latter is a higher-rank Maxwell-like action \cite{Bertolini:2022ijb}, whose Equations of Motion (EoM) represent a rank-2 generalization of the standard electromagnetism which describes fractonic phenomena \cite{Pretko:2016lgv,Bertolini:2022ijb}. In 3D the theory appears to be a traceless non-topological rank-2 generalization of the ordinary Chern-Simons model \cite{Bertolini:2024yur}, which exhibits a fractonic Hall-like behaviour, characterized by a dipole-flux attachment and a generalized Hall current.\\
 
Thinking about topological theories and their relations to condensed matter, the higher-rank covariant generalization of the Chern-Simons case \cite{Bertolini:2024yur} represents a first promising step towards the study of 3D covariant fracton models. Inspired by that successful example, in this work we build and study the covariant rank-2 generalization of another important 3D topological field theory: the BF model \cite{Birmingham:1991ty}. The aim is to investigate whether fractonic excitations naturally emerge from a pure field theoretical perspective, and eventually establish a link with the noncovariant model \cite{Ebisu:2023idd} and relations with the R2TC. We therefore first study and characterize the physical content of the covariant rank-2 BF theory, and, once matter is introduced, we show that subdimensional quasiparticles emerge, such as fractons and lineons. Interestingly, our continuum theory can be mapped to the dipolar BF theory discussed in \cite{Oh:2022klh, Han:2024nvu}, related to the rank-2 $U(1)$ gauge theory, which is the effective field description of the R2TC. Finally, by analyzing the case of two completely symmetric tensor gauge fields, we obtain a rank-2 generalization of the action which describes the low energy physics of topological insulators \cite{Cho:2010rk}, relevant in the context of topological dipole insulators, as recently proposed in \cite{Lam:2024smz}.\\  
 
The paper is organized as follows. {In Section \ref{sec-model} the action of the model is identified by locality, power counting and invariance under the symmetry transformations. We also analyze the energy-momentum tensor which has the peculiar property of vanishing on-shell, thus making the model almost topological. In Section \ref{sec-propagator} we introduce a general covariant gauge fixing term which is needed to compute the propagators. The details are in Appendix \ref{App:propagators}.} In Section \ref{sec-dof} the DoF of the theory are counted. In Section \ref{sec-currents} a coupling to matter is introduced  and a subdimensional behaviour is observed, studied and connected to the pre-existing condensed matter literature. In Section \ref{Sec-symm} the completely symmetric case is investigated and a fractonic behaviour emerges when the theory is coupled to matter. In Section \ref{sec-conclusion} we write our conclusions.

\subsection*{Notation and conventions}

Spacetime dimension: $3D=2+1; 4D=3+1$.\\
Greek indices:  $\mu,\nu,\rho,...=0,1,2$.\\
Latin  indices $i,j,k,...=1,2.$\\
Minkowski metric: $ \eta_{\mu\nu}=\mbox{diag}(-1,1,1)$. \\
Levi-Civita symbol: $\epsilon_{012}\equiv 1= -\epsilon^{012}$.

\section{The model}\label{sec-model}

\subsection{Symmetries and equations of motion}

The field content of the theory consists of  two rank-2 tensor fields: $a_{\mu\nu}(x)$, which is symmetric $a_{\mu\nu}(x)=a_{\nu\mu}(x)$,  and $B_{\mu\nu}(x)$, which has no symmetry. Restricting to functionals with one derivative only, the most general 3D actions depending on these two tensor fields and invariant under the following field transformations
	\bea
		\delta_1 a_{\mu\nu} = \partial_{\mu}\partial_{\nu}\Lambda\quad &;& \quad 
		\delta_1 B_{\mu\nu} = 0  \label{gauge transformation 1}\\
		\delta_2 a_{\mu\nu} = 0 \quad &;& \quad 
		\delta_2 B_{\mu\nu} =\partial_\mu \xi_\nu\ , \label{gauge transformation 2}
\eea
are
\bea
S_{CS}^{(a)} &=&  \int d^3 x\, \epsilon^{\mu\nu\rho} a_\mu^{\ \lambda}\partial_\nu a_{\rho\lambda} 
\label{csa}\\
S_{CS}^{(B)}  &=&  \int d^3 x\, \epsilon^{\mu\nu\rho} B_\mu^{\ \lambda}\partial_\nu B_{\rho\lambda} 
\label{csB}\\
S_{BF} &=& \int d^3 x\, \epsilon^{\mu\nu\rho}\invmixt{B}{\mu}{\sigma}\partial_\nu a_{\rho\sigma}\label{Sinv}\ ,
\eea
where the tensor fields have mass dimensions $[a_{\mu\nu}]=[B_{\mu\nu}]=1.$ 
The $\delta_1$-transformation acting on the symmetric tensor field $a_{\mu\nu}(x)$ represents the longitudinal diffeomorphisms characterizing the covariant formulation of fracton theories both in 4D \cite{Bertolini:2022ijb,Bertolini:2023juh,Bertolini:2023sqa} and 3D \cite{Bertolini:2024yur}. On the other hand, the $\delta_2$-transformation is the lowest order most general one acting on the generic tensor field $B_{\mu\nu}(x)$. It reduces to infinitesimal diffeomorphisms (which contain their longitudinal component) in the particular case of symmetric $B_{\mu\nu}(x)$, and, for antisymmetric tensor field, 
$\delta_2 B_{\mu\nu} =\frac{1}{2}(\partial_\mu \xi_\nu - \partial_\nu \xi_\mu)$ is the standard transformation of the 2-form appearing in ordinary BF models \cite{Birmingham:1991ty}. This transformation has been shown to display fractonic behaviours as well \cite{Bertolini:2025qcy}. 
The first two actions, \eqref{csa} and \eqref{csB}, are the higher-rank extension of the ordinary Chern-Simons action for the vector gauge field $A_\mu(x)$
\be
{S}^{(ord)}_{CS} =  \int d^3 x\, \epsilon^{\mu\nu\rho} A_\mu\partial_\nu A_\rho\ .
\label{cs}\ee
In particular, the higher-rank Chern-Simons action \eqref{csa} has been studied in \cite{Bertolini:2024yur}, where it has been shown that it describes fracton quasiparticles exhibiting a Hall-like behaviour. 
The last action functional \eqref{Sinv} is the generalization of the 3D BF action \cite{Fradkin:2013sab, Birmingham:1991ty}, which couples the vector gauge field $A_\mu(x)$ to an additional gauge field $B_\mu(x)$
\be
{S}^{(ord)}_{BF} =  \int d^3 x\, \epsilon^{\mu\nu\rho} B_\mu\partial_\nu A_\rho\ .
\label{bf}\ee
In this paper we focus on the action $S_{BF}$ \eqref{Sinv}, which can be isolated by means of a discrete symmetry which assigns opposite charges to the fields $a_{\mu\nu}(x)$ and $B_{\mu\nu}(x)$. For instance, definining
\bea
{\cal P} a_{\mu\nu} &=& +\, a_{\mu\nu} \\
{\cal P} B_{\mu\nu}&=& -\, B_{\mu\nu}\ ,
\eea
the charges of the actions \eqref{csa}, \eqref{csB} and \eqref{Sinv} are
\bea
{\cal P} S_{CS}^{a} &=& +2\, S_{CS}^{a} \\
{\cal P} S_{CS}^{B}&=& - 2\,S_{CS}^{B} \\
{\cal P} S_{BF} &=& 0\ .\label{Pcharge}
\eea
Requiring that the action has vanishing ${\cal P}$-charge uniquely identifies $S_{BF}$ \eqref{Sinv}.
The action \eqref{Sinv} can be written in a form that is even more reminiscent to the standard BF action
	\be
	S_{BF} = \frac{1}{3}\int d^3 x\,\epsilon^{\mu\nu\rho}\invmixt{B}{\mu}{\sigma}F_{\sigma\nu\rho}\ ,\label{S_BF}
	\ee
where we used the higher-rank field strength introduced in \cite{Bertolini:2022ijb}
	\begin{equation}
	F_{\mu\nu\rho} \equiv \partial_\mu a_{\nu\rho} + \partial_\nu a_{\mu\rho} -2 \partial_\rho a_{\mu\nu}\ ,\label{def_fracton_field_strength}
	\end{equation}
which is invariant under the field transformations \eqref{gauge transformation 1} and \eqref{gauge transformation 2}
\be
	\delta_{1} F_{\mu\nu\rho}=\delta_{2} F_{\mu\nu\rho}=0\ ,
\ee
and whose properties are
	\begin{equation}
	F_{\mu\nu\rho}=F_{\nu\mu\rho}
	\end{equation}
	\begin{equation}\label{Bianchi 3D fractons}
	\epsilon_{\nu\rho\sigma}\partial^\nu F^{\mu\rho\sigma}=0
	\end{equation}
	\be\label{cyclicity field strength}
	F_{\mu\nu\rho} + F_{\rho\mu\nu} + F_{\nu\rho\mu}=0\,.
	\ee
It is useful to decompose the generic tensor $B_{\mu\nu}(x)$ in its symmetric and antisymmetric parts
	\begin{equation}
	 B_{\mu\nu} \equiv b_{\mu\nu}+ \epsilon_{\mu\nu\rho}b^\rho\ ,\label{decomposition_B}
	\end{equation}
where
	\begin{align}
	b_{\mu\nu}=b_{\nu\mu}&=\tfrac{1}{2}\left(B_{\mu\nu}+B_{\nu\mu}\right)\\
	\epsilon_{\mu\nu\rho}b^\rho&=\tfrac{1}{2}\left(B_{\mu\nu}-B_{\nu\mu}\right)
	\rightarrow
	b^\rho=-\tfrac{1}{2}\epsilon^{\rho\alpha\beta}B_{\alpha\beta}\ ,
	\end{align}
which, from \eqref{gauge transformation 2}, transform as
	\begin{align}
	\delta_2 b_{\mu\nu}&=\tfrac{1}{2}\left(\partial_\mu\xi_\nu+\partial_\nu\xi_\mu\right)\label{gauge transformation 2_symmetric_part}\\
	\delta_2b^\rho&=-\tfrac{1}{2}\epsilon^{\rho\alpha\beta}\partial_\alpha\xi_\beta\ .
	\end{align}
The invariant action \eqref{Sinv}, written  in terms of the fields $a_{\mu\nu}(x)$, $b_{\mu\nu}(x)$ and $b_\mu(x)$, reads
	\be
	S_{BF}
	= \int d^3x\left(
	\epsilon^{\mu\nu\rho}\invmixt{b}{\mu}{\sigma}\partial_\nu a_{\rho\sigma}-b^\mu \partial^\nu a_{\mu\nu}+b^\mu\partial_\mu a\right)\ ,
	\ee
where the trace $a(x)$ is defined by means of the Minkowski metric
	\begin{equation}
	a\equiv \eta^{\mu\nu}a_{\mu\nu}\ .
	\end{equation}
The gauge transformations $\delta_1$ \eqref{gauge transformation 1} and $\delta_2$ \eqref{gauge transformation 2} depend on the scalar gauge parameter $\Lambda(x)$ and on the vector $\xi_\mu(x)$, respectively, and, correspondingly, require a scalar and a vector gauge condition
	\begin{align}
	{k_0} \partial^\mu \partial^\nu a_{\mu\nu} + {k_1} \partial^2 a&=0 \label{scalar gauge fixing}\\
	\kappa_0 \partial^\nu b_{\mu\nu}+ \kappa_1 \partial_\mu b+ \kappa_2 \epsilon_{\mu\nu\rho}\partial^\nu b^\rho  &=0\ , \label{vector gauge fixing}
	\end{align}
where 
	\begin{equation}
	b\equiv \eta^{\mu\nu}b_{\mu\nu}\ ,
	\end{equation}
and $k_0$, $k_1$, $\kappa_0$, $\kappa_1$ and $\kappa_2$ are constant parameters. The above gauge conditions
can be implemented by adding to the invariant action $S_{BF}$ \eqref{Sinv} the gauge fixing term
	\be
	S_{gf}=\int d^3x\left[d\left(k_0\partial^\mu\partial^\nu a_{\mu\nu}+k_1\partial^2 a+\frac{k}{2} d\right)+d^\mu\left(\kappa_0\partial^\nu b_{\mu\nu}+\kappa_1\partial_\mu b+\kappa_2\epsilon_{\mu\nu\rho}\partial^\nu b^\rho+\frac{\kappa}{2} d_\mu\right)\right]\ ,\label{S_gf_multipliers}
	\ee
where $d(x)$ and $d^\mu(x)$ are Nakanishi-Lautrup multipliers \cite{Nakanishi:1966zz,Lautrup:1967zz}, with mass dimensions
\begin{align}
    [d]=0 \qquad
     [d_\mu]=1\ ,
\end{align}
which imply that the gauge parameters $k$ and $\kappa$ have non-vanishing mass dimensions
	\be
	[{k}]=3 \qquad
	[\kappa ]=1\ ,
	\ee
which renders the Landau gauge mandatory, in order to avoid infrared divergences \cite{Birmingham:1991ty,Alvarez-Gaume:1989ldl}
	\be\label{Landau}
	{k}=\kappa =0\ .
	\ee
The EoM of the gauge fixed action
	\be
	S\equiv S_{BF} + S_{gf}\ , \label{def gauge fixed action}
	\ee
are
	\begin{align}
	\frac{\delta S}{\delta a_{\alpha\beta}}&=\frac{1}{2}\left(\epsilon^{\alpha\mu\nu}\partial_\mu b_\nu^{\ \beta}+\epsilon^{\beta\mu\nu}\partial_\mu b_\nu^{\ \alpha}+\partial^\alpha b^\beta+\partial^\beta b^\alpha\right)-\eta^{\alpha\beta}\partial_\mu b^\mu+k_0\partial^\alpha\partial^\beta d+k_1\eta^{\alpha\beta}\partial^2d\label{E-eoma-gf}\\
	\frac{\delta S}{\delta b_{\alpha\beta}}&=\frac{1}{2}\left(\epsilon^{\alpha\mu\nu}\partial_\mu a_\nu^{\ \beta}+\epsilon^{\beta\mu\nu}\partial_\mu a_\nu^{\ \alpha}\right)-\frac{\kappa_0}{2}\left(\partial^\alpha d^\beta+\partial^\beta d^\alpha\right)-\kappa_1\eta^{\alpha\beta}\partial_\mu d^\mu\label{E-eomh-gf}\\
	\frac{\delta S}{\delta b^\alpha}&=-\partial^\mu a_{\mu\alpha}+\partial_\alpha a+\kappa_2\epsilon_{\alpha\mu\nu}\partial^\mu d^\nu\label{E-eomv-gf}\\
	\frac{\delta S}{\delta d}&=k_0\partial^\mu\partial^\nu a_{\mu\nu}+k_1\partial^2a\\
	\frac{\delta S}{\delta d^\alpha}&=\kappa_0\partial^\mu b_{\mu\alpha}+\kappa_1\partial_\alpha b+\kappa_2\epsilon_{\alpha\mu\nu}\partial^\mu b^\nu\ .\label{E-eomd-gf}
	\end{align}
From \eqref{E-eoma-gf} and \eqref{E-eomh-gf} we get
	\begin{align}
	\eta_{\alpha\beta}\fdv{S}{a_{\alpha\beta}} &= -2\partial_\mu b^\mu + ({k_0} + 3{k_1})\partial^2 d\\
	\eta_{\alpha\beta}\fdv{S}{b_{\alpha\beta}} &= -(\kappa_0 + 3\kappa_1)\partial_\mu d^\mu\ ,
	\end{align}
so we see that,  if 
\be
{k_0} + 3{k_1}=0\ ,
\ee
the trace $a(x)$ plays the role of a multiplier for the condition
\be
\partial^\mu b_\mu =0\ ,
\label{db=0}\ee
while $S$ depends on the trace $b(x)$ only through the gauge fixing term, unless 
\be
\kappa_0 + 3\kappa_1=0\ ,
\label{k0+3k1}\ee
in which case $b(x)$ disappears completely from the gauge fixed action $S$ \eqref{def gauge fixed action}, which renders \eqref{k0+3k1} a naturally preferred choice.

\subsection{Energy-momentum tensor}

Making explicit the dependence of the invariant action $S_{BF}$ \eqref{Sinv}
on the metric in a curved spacetime
\be
S_{BF}= \int \dd^3 x\, \epsilon^{\mu\nu\rho}g^{\sigma\lambda}\invmixt{B}{\mu\lambda}{}\partial_\nu a_{\rho\sigma}\ ,
\label{Sinvcurved}\ee
we can compute the energy-momentum tensor
\be
T_{\alpha\beta} \equiv -\frac{2}{\sqrt{-g}}\fdv{S_{BF}}{g^{\alpha\beta}}\bigg|_{g=\eta}
= -B_{\mu\alpha}\epsilon^{\mu\nu\rho}\partial_\nu a_{\rho\beta}-B_{\mu\beta}\epsilon^{\mu\nu\rho}\partial_\nu a_{\rho\alpha}\ .
\label{E-m_tensor}\ee
Using the EoM, written in terms of $B_{\mu\nu}(x)$ and $a_{\mu\nu}(x)$ 
\bea
    \fdv{S_{BF}}{B_{\alpha\beta}} &=& \epsilon^{\alpha\nu\rho}\partial_\nu \invmixt{a}{\rho}{\beta}
    \label{EoMB}\\
    \fdv{S_{BF}}{a_{\alpha\beta}} &=& \frac{1}{2}\left(
    \epsilon^{\alpha\mu\nu}\partial_\mu \invmixt{B}{\nu}{\beta}+\epsilon^{\beta\mu\nu}\partial_\mu \invmixt{B}{\nu}{\alpha}
    \right)\ ,
    \label{EoMd}
\eea
it is readily seen that on-shell the energy-momentum tensor vanishes:
\be
    \left.T_{\alpha\beta}\right |_{\mbox{\tiny on-shell}} =0\ .
\ee
This is analogous to topological quantum field theories \cite{Birmingham:1991ty}, where the only contribution to the energy-momentum tensor comes from the gauge fixing term. The difference is that in topological quantum field theories the contribution to $T_{\mu\nu}(x)$ from $S_{BF}$  \eqref{Sinv} vanishes off-shell, 
which is a stronger property than the on-shell vanishing we are observing in this case. This latter is quite a peculiar feature of this theory, which is not topological, because on a curved manifold the action $S_{BF}$ does depend on the generic spacetime metric $g_{\mu\nu}(x)$, although the dependence is mild, as it is apparent if we compare the action \eqref{Sinvcurved} with, for instance, 3D Maxwell theory
\be
S_{Max} = -\frac{1}{4} \int d^3 x\, \sqrt{|g|}g^{\mu\rho}g^{\nu\sigma}F_{\mu\nu}F_{\rho\sigma}
 \ ,
\label{Smaxcurved}\ee
which allows us to call the theory described by the action $S_{BF}$ \eqref{Sinv} as ``quasi-topological''.

\section{Propagators}\label{sec-propagator}

The gauge fixed action \eqref{def gauge fixed action} in the Landau gauge \eqref{Landau} and  in momentum space\footnote{The Fourier transform is defined as $\Phi(x) \equiv \int \frac{\dd^3 p}{(2\pi)^3} e^{i p\cdot x} \hat{\Phi}(p)$. }, writes 
	\begin{align}
	S&=\int \frac{d^3 p}{(2\pi)^3}\left\{i\hat a_{\mu\nu}(-p)\left[ \epsilon^{\mu\lambda\rho}p_\lambda\hat{b}_\rho^{\ \nu}(p)+p^\mu \hat b^\nu(p)-\eta^{\mu\nu}p_\lambda \hat b^\lambda(p)\right]-\hat d(-p)\left(k_0p^\alpha p^\beta+k_1p^2\eta^{\alpha\beta}\right)\hat a_{\alpha\beta}(p)+\right.\nonumber\\
	&\qquad\qquad\ \left.+i\hat d^\mu(-p)\left[\kappa_0p^\nu\hat{b}_{\mu\nu}(p)+\kappa_1\eta^{\alpha\beta}p_\mu\hat{b}_{\alpha\beta}(p)+\kappa_2\epsilon_{\mu\nu\rho}p^\nu \hat b^\rho(p)\right]\right\}\nonumber\\
	&\equiv\int \frac{d^3 p}{(2\pi)^3}\left[\hat a_{\mu\nu}(-p)G^{\mu\nu,\alpha\beta}\hat{b}_{\alpha\beta}(p)+\hat a_{\mu\nu}(-p)G^{\mu\nu,\alpha}\hat b_\alpha(p)+\hat d(-p)G^{\alpha\beta}\hat a_{\alpha\beta}(p)+\right.\nonumber\\
	&\qquad\qquad\ \left.+\hat d_\mu(-p)G^{\alpha\beta,\mu}_{^{(\kappa_0,\kappa_1)}}\hat{b}_{\alpha\beta}(p)+\hat d_\mu(-p)G^{\mu,\alpha}\hat b_\alpha(p)\right]\nonumber\\
	&\equiv\int \frac{d^3 p}{(2\pi)^3}\,\hat\phi_M(-p)G^{MA}\hat\phi_A(p)\ ,
	\end{align}
where we defined
	\begin{align}
	\hat\phi_M&\equiv\{\hat a_{\mu\nu}\ ,\ \hat{b}_{\mu\nu}\ ,\ \hat b_\mu\ ,\ \hat d_\mu\ ,\ \hat d\}\ ,\\
	G^{MA}&\equiv\frac{1}{2}\left[
		\begin{array}{ccccc}
		0&G^{\mu\nu,\alpha\beta}&  G^{\mu\nu,\alpha}&0&  G^{\mu\nu}\\
		G^{\mu\nu,\alpha\beta}&0&0&G^{*\mu\nu,\alpha}_{^{(\kappa_0,\kappa_1)}}&0\\
		G^{*\alpha\beta,\mu}&0&0&G^{\mu,\alpha}&0\\
		0&G^{\alpha\beta,\mu}_{^{(\kappa_0,\kappa_1)}}&G^{*\alpha,\mu}&0&0\\
		G^{\alpha\beta}&0&0&0&0
		\end{array}\right]\ ,\label{GAB}
	\end{align}
and
	\begin{align}
	G^{\mu\nu,\alpha\beta} (p)&\equiv
	\tfrac{i}{4}p_\lambda (
	\epsilon^{\mu\lambda\alpha} \eta^{\beta\nu} +
	\epsilon^{\nu\lambda\alpha} \eta^{\beta\mu} +
	\epsilon^{\mu\lambda\beta} \eta^{\alpha\nu} +
	\epsilon^{\nu\lambda\beta} \eta^{\alpha\mu}
	)
	\label{E-A5'}\\
	G^{\mu\nu,\alpha}(p)&\equiv\tfrac i 2\left(\eta^{\mu\alpha}p^\beta+\eta^{\mu\beta}p^\alpha\right)-ip^\mu\eta^{\alpha\beta}\\
	G^{\mu\nu}(p)&\equiv-k_0p^\mu p^\nu-k_1p^2\eta^{\mu\nu}\\
	G^{\mu\nu,\alpha}_{^{(\kappa_0,\kappa_1)}}(p)&\equiv\tfrac i 2 \kappa_0\left(\eta^{\mu\alpha}p^\nu+\eta^{\nu\alpha}p^\mu\right)+i\kappa_1p^\alpha\eta^{\mu\nu}\\
	G^{\mu,\alpha}(p)&\equiv i\kappa_2\epsilon^{\mu\lambda\alpha}p_\lambda\ ,
	\end{align}
which display the following symmetries
	\begin{align}
	G^{\mu\nu,\alpha\beta} &=G^{*\alpha\beta,\mu\nu}=G^{\nu\mu,\alpha\beta}=G^{\mu\nu,\beta\alpha}\label{symm_of_G}\\
	G^{\mu\nu}&=G^{\nu\mu}=G^{*\mu\nu}\\
	G^{\mu\nu,\alpha}_{^{(\kappa_0,\kappa_1)}}&=G^{\nu\mu,\alpha}_{^{(\kappa_0,\kappa_1)}}=-G^{*\mu\nu,\alpha}_{^{(\kappa_0,\kappa_1)}}\label{GK}\\
	G^{\mu,\alpha}&=-G^{*\mu,\alpha}=G^{*\alpha,\mu}\ .
	\end{align}	
The propagators of the theory are encoded in the matrix
\be\label{Delta}
	\Delta_{AP}\equiv\left[
		\begin{array}{ccccc}
		\Delta^{_{(1)}}_{\alpha\beta,\rho\sigma}&\Delta^{_{(2)}}_{\alpha\beta,\rho\sigma}&\Delta^{_{(3)}}_{\alpha\beta,\rho}&\Delta^{_{(4)}}_{\alpha\beta,\rho}&\Delta^{_{(5)}}_{\alpha\beta}\\
		\Delta^{_{(2)}*}_{\rho\sigma,\alpha\beta}&\Delta^{_{(6)}}_{\alpha\beta,\rho\sigma}&\Delta^{_{(7)}}_{\alpha\beta,\rho}&\Delta^{_{(8)}}_{\alpha\beta,\rho}&\Delta^{_{(9)}}_{\alpha\beta}\\
		\Delta^{_{(3)}*}_{\rho\sigma,\alpha}&\Delta^{_{(7)}*}_{\rho\sigma,\alpha}&\Delta^{_{(10)}}_{\alpha,\rho}&\Delta^{_{(11)}}_{\alpha,\rho}&\Delta^{_{(12)}}_\alpha\\
		\Delta^{_{(4)}*}_{\rho\sigma,\alpha}&\Delta^{_{(8)}*}_{\rho\sigma,\alpha}&\Delta^{_{(11)}*}_{\rho,\alpha}&\Delta^{_{(13)}}_{\alpha,\rho}&\Delta^{_{(14)}}_\alpha\\
		\Delta^{_{(5)}*}_{\rho\sigma}&\Delta^{_{(9)}*}_{\rho\sigma}&\Delta^{_{(12)}*}_\rho&\Delta^{_{(14)}*}_\rho&\Delta^{_{(15)}}
		\end{array}\right]\ ,
	\ee
with
\begin{align}
\Delta^{_{(1)}}_{\alpha\beta,\rho\sigma}(p)&\equiv \langle \hat a_{\alpha\beta}(-p)\,\hat  a_{\rho\sigma}(p)\rangle&
\Delta^{_{(2)}}_{\alpha\beta,\rho\sigma}(p)&\equiv \langle \hat a_{\alpha\beta}(-p)\, \hat{b}_{\rho\sigma}(p)\rangle\label{E-<ah>}\\
\Delta^{_{(3)}}_{\alpha\beta,\rho}(p)&\equiv \langle \hat a_{\alpha\beta}(-p)\, \hat b_{\rho}(p)\rangle&
\Delta^{_{(4)}}_{\alpha\beta,\rho}(p)&\equiv \langle \hat a_{\alpha\beta}(-p)\, \hat d_{\rho}(p)\rangle\\
\Delta^{_{(5)}}_{\alpha\beta}(p)&\equiv \langle \hat a_{\alpha\beta}(-p)\, \hat d(p)\rangle&
\Delta^{_{(6)}}_{\alpha\beta,\rho\sigma}(p)&\equiv \langle \hat{b}_{\alpha\beta}(-p)\, \hat{b}_{\rho\sigma}(p)\rangle\\
\Delta^{_{(7)}}_{\alpha\beta,\rho}(p)&\equiv \langle \hat{b}_{\alpha\beta}(-p)\, \hat b_{\rho}(p)\rangle&
\Delta^{_{(8)}}_{\alpha\beta,\rho}(p)&\equiv \langle \hat{b}_{\alpha\beta}(-p)\, \hat d_{\rho}(p)\rangle\label{E-<bh>}\\
\Delta^{_{(9)}}_{\alpha\beta}(p)& \equiv\langle \hat{b}_{\alpha\beta}(-p)\,\hat d(p)\rangle&
\Delta^{_{(10)}}_{\alpha,\rho}(p)&\equiv \langle \hat b_{\alpha}(-p)\,\hat b_{\rho}(p)\rangle\\
\Delta^{_{(11)}}_{\alpha,\rho}(p)&\equiv  \langle \hat b_{\alpha}(-p)\, \hat d_{\rho}(p)\rangle&
\Delta^{_{(12)}}_\alpha(p)&\equiv  \langle \hat b_{\alpha}(-p)\, \hat d(p)\rangle\\
\Delta^{_{(13)}}_{\alpha,\rho}(p)&\equiv  \langle \hat d_{\alpha}(-p)\, \hat d_{\rho}(p)\rangle&
\Delta^{_{(14)}}_\alpha(p)&\equiv \langle \hat d_{\alpha}(-p)\, \hat d(p)\rangle\\
\Delta^{_{(15)}}(p)&\equiv \langle \hat d(-p)\, \hat d(p)\rangle\ ,
\end{align}
where
	\be
	\Delta_{AP}=\Delta^\dag_{PA}\ ,
	\ee
and the symmetries
	\be
	\Delta^{_{(i)}}_{M,P}(p)=\Delta^{_{(i)}*}_{P,M}(p)=\Delta^{_{(i)}*}_{M,P}(-p) \quad\mbox{for}\ i=\{1,6,10,11,13\}
	\ee
$i.e.$ when $M=\mu\nu$ and $P=\rho\sigma$ or $M=\mu$ and $P=\rho$, and
	\be
	\Delta^{_{(i)}}_{\alpha\beta,P}(p)=\Delta^{_{(i)}}_{\beta\alpha,P}(p)=\Delta^{_{(i)}*}_{\alpha\beta,P}(-p)\quad\mbox{for}\ i=\{1,...,9\}
	\ee
$i.e.$ when $P=\{\rho\sigma\ ,\ \sigma\rho\ ,\ \rho\ ,\ ...\ \}$, have been taken into account. The propagator matrix $\Delta_{AP}(p)$ \eqref{Delta} is defined as the inverse of the quadratic operator $G^{AB}(p)$ \eqref{GAB}
	\be\label{E-KD=I}
	G^{MA}\Delta_{AP}=I^M_P\ ,
	\ee
where
	\be
	I^M_{P}\equiv\left[
		\begin{array}{ccccc}
		\mathcal{I}^{\mu\nu}_{\rho\sigma}(0)&0&0&0&0\\
		0&\mathcal{I}^{\mu\nu}_{\rho\sigma}(\lambda)&0&0&0\\
		0&0&\delta^\mu_\rho&0&0\\
		0&0&0&\delta^\mu_{\rho}&0\\
		0&0&0&0&1
		\end{array}\right]
	\ee
is the identity matrix with
	\be
	\mathcal{I}^{\mu\nu}_{\rho\sigma}(\lambda)\equiv \frac{1}{2}\left(
	\delta^\mu_\rho \delta^\nu_\sigma + \delta^\mu_\sigma \delta^\nu_\rho
	\right)+\lambda\eta^{\mu\nu}\eta_{\rho\sigma}\label{def cal_I(lambda)}\,.
	\ee
A comment is in order concerning the presence of the constant $\lambda$ in the matrix identity. The idempotency of the identity
	\be
	\mathcal{I}^{\mu\nu}_{\rho\sigma}\,\mathcal{I}^{\rho\sigma}_{\alpha\beta} = \mathcal{I}^{\mu\nu}_{\alpha\beta}\ ,
	\ee
requires that either 
	\begin{align}
	&\lambda=0\ ,\\
	\mbox{or}&\nonumber\\
	&\lambda= -\frac{1}{3}\, \label{a=-1/3}.
	\end{align}
The latter option \eqref{a=-1/3} corresponds to a traceless identity
\begin{equation}
    \eta_{\mu\nu}\mathcal{I}^{\mu\nu}_{\rho\sigma}(-\tfrac{1}{3})=\eta^{\rho\sigma}\mathcal{I}^{\mu\nu}_{\rho\sigma}(-\tfrac{1}{3})=0\ ,
\end{equation}
which is suitable for a traceless tensorial space such as that involving the field $b_{\mu\nu}(x)$. It is instructive that, leaving it as a free parameter, we find that the propagators are defined only when $\lambda=-\frac{1}{3}$ and 
\begin{equation}
    \kappa_0 + 3\kappa_1 =0\ ,
\end{equation}
\textit{i.e.} when the gauge fixed action \eqref{def gauge fixed action} 
depends only on the traceless part of $B_{\mu\nu}(x)$~:
	\be
	\tilde{B}_{\mu\nu}\equiv B_{\mu\nu}-\frac{1}{3}\eta_{\mu\nu}b\,.
	\label{Btraceless}\ee
For the same reason, we expect that the propagators involving the trace $b(x)$ will be vanishing. We choose not to impose this constraint, although reasonable, to leave the tracelessness property as a check of the correctness of our approach.   Indeed solutions to the system of equations given by \eqref{E-KD=I}, computed in Appendix \ref{App:propagators}, exist only if
	\be\label{E-K1}
	\lambda=-\frac{1}{3}\quad;\quad\kappa_0+3\kappa_1=0\ ,
	\ee
and give the following non-trivial propagators (see Appendix A for the details of the calculation)
	\begin{align}
	\Delta^{_{(2)}}_{\alpha\beta,\rho\sigma}(p)&\equiv \langle \hat a_{\alpha\beta}(-p)\, \hat{b}_{\rho\sigma}(p)\rangle\label{Delta2}\\
&=	-\frac{i}{2p^2}p^\lambda\bigg[\left(	\epsilon_{\alpha\lambda\rho}  t_{\sigma\beta} +	\epsilon_{\beta\lambda\rho}  t_{\sigma\alpha} +	\epsilon_{\alpha\lambda\sigma}  t_{\rho\beta} +	\epsilon_{\beta\lambda\sigma}  t_{\rho\alpha}\right)+\nonumber\\
	&\quad-\frac{4\kappa_2}{\kappa_0-\kappa_2}	  \left(\epsilon_{\alpha\lambda\rho}\tfrac{p_\sigma p_\beta}{p^2}  +\epsilon_{\alpha\lambda\sigma} \tfrac{p_\rho p_\beta}{p^2} +	\epsilon_{\beta\lambda\rho}\tfrac{ p_\sigma p_\alpha}{p^2} +\epsilon_{\beta\lambda\sigma} \tfrac{p_\rho p_\alpha}{p^2}\right)\bigg]\nonumber\\[5px]
	\Delta^{_{(3)}}_{\alpha\beta,\rho}(p)&\equiv \langle \hat a_{\alpha\beta}(-p)\, \hat b_{\rho}(p)\rangle\label{Delta3}\\
	&= \frac{i}{p^2}\left[\frac{2\kappa_0}{\kappa_0-\kappa_2}\left(t_{\alpha\rho}p_\beta+t_{\beta\rho}p_\alpha\right)-\frac{1}{k_0+k_1}\left(k_0\, t_{\alpha\beta}p_\rho+k_1\, \tilde t_{\alpha\beta}p_\rho\right)\right]\nonumber\\[5px]
	\Delta^{_{(5)}}_{\alpha\beta}(p)&\equiv \langle \hat a_{\alpha\beta}(-p)\, \hat d(p)\rangle= -\frac{2}{k_0+k_1}\frac{p_\alpha p_\beta}{p^4}\label{Delta5}\\[5px]
	\Delta^{_{(8)}}_{\alpha\beta,\rho}(p)&\equiv \langle \hat{b}_{\alpha\beta}(-p)\, \hat d_{\rho}(p)\rangle= \frac{i}{p^2}\left[\frac{1}{\kappa_0}\tilde t_{\alpha\beta}p_\rho-\frac{2}{\kappa_0-\kappa_2}\left(t_{\alpha\rho}p_\beta+t_{\beta\rho}p_\alpha\right)\right]\label{Delta8}\\[5px]
	\Delta^{_{(11)}}_{\alpha,\rho}(p)&\equiv  \langle \hat b_{\alpha}(-p)\, \hat d_{\rho}(p)\rangle=\frac{2i}{\kappa_0-\kappa_2}\frac{\epsilon_{\alpha\lambda\rho}p^\lambda}{p^2}\ ,\label{Delta11}
	\end{align}
where
	\be
	t_{\alpha\beta}\equiv \eta_{\alpha\beta}-\tfrac{p_\alpha p_\beta}{p^2}\quad;\quad\tilde t_{\alpha\beta}\equiv \eta_{\alpha\beta}-3\tfrac{p_\alpha p_\beta}{p^2}\ ,\label{def_t_tildet}
	\ee
such that
	\be
	p^\alpha t_{\alpha\beta}=0\quad;\quad\eta^{\alpha\beta}\tilde t_{\alpha\beta}=0\,. \label{traces_prop}
	\ee
It is thus immediate to observe that
	\be
	\eta^{\rho\sigma}\Delta^{_{(2)}}_{\alpha\beta,\rho\sigma}(p)=\langle \hat a_{\alpha\beta}(-p)\, \hat{b}(p)\rangle=0=\eta^{\alpha\beta}\Delta^{_{(8)}}_{\alpha\beta,\rho}(p)=\langle \hat{b}(-p)\, \hat d_{\rho}(p)\rangle\ ,
	\ee
which confirms that the trace $b(x)$ has no role in the theory. The propagators \eqref{Delta2}-\eqref{Delta11} display poles for
	\be\label{E-poles}
	\kappa_0=0\quad;\quad\kappa_0=\kappa_2\quad;\quad k_1=-k_0\ ,
	\ee
which identify values of the gauge fixing parameters for which the propagators of the theory, hence the theory itself, are not defined.

\section{Degrees of Freedom}\label{sec-dof}

The on-shell EoM \eqref{E-eoma-gf}-\eqref{E-eomd-gf} in momentum space read
	\begin{align}
	\frac{\delta S}{\delta \hat a_{\alpha\beta}}&=\frac{i}{2}\left(\epsilon^{\alpha\mu\nu} p_\mu \hat{b}_\nu^{\ \beta}+\epsilon^{\beta\mu\nu} p_\mu \hat{b}_\nu^{\ \alpha}+ p^\alpha \hat b^\beta+ p^\beta \hat b^\alpha\right)-i\eta^{\alpha\beta} p_\mu \hat b^\mu-k_0 p^\alpha p^\beta \hat d-k_1\eta^{\alpha\beta} p^2\hat d=0\label{E-eomPa-gf}\\
	\frac{\delta S}{\delta \hat{b}_{\alpha\beta}}&=\frac{1}{2}\left(\epsilon^{\alpha\mu\nu} p_\mu \hat a_\nu^{\ \beta}+\epsilon^{\beta\mu\nu} p_\mu \hat a_\nu^{\ \alpha}\right)-\frac{\kappa_0}{2}\left( p^\alpha \hat d^\beta+ p^\beta \hat d^\alpha\right)+\frac{1}{3}\kappa_0\eta^{\alpha\beta} p_\mu \hat d^\mu=0\label{E-eomPh-gf}\\
	\frac{\delta S}{\delta \hat b^\alpha}&=- p^\mu \hat a_{\mu\alpha}+ p_\alpha \hat a+\kappa_2\epsilon_{\alpha\mu\nu} p^\mu \hat d^\nu=0\label{E-eomPv-gf}\\
	\frac{\delta S}{\delta \hat d}&=k_0 p^\mu p^\nu \hat a_{\mu\nu}+k_1 p^2\hat a=0\label{E-eomPd}\\
	\frac{\delta S}{\delta \hat d^\alpha}&=\kappa_0 p^\mu \hat{b}_{\mu\alpha}-\frac{1}{3}\kappa_0 p_\alpha \hat{b}+\kappa_2\epsilon_{\alpha\mu\nu} p^\mu \hat b^\nu=\kappa_0 p^\mu \hat{\tilde{b}}_{\alpha\mu}+\kappa_2 \epsilon_{\alpha\mu\nu}p^\mu \hat b^\nu
=0\ ,\label{E-eomPBvec}
	\end{align}
where the tracelessness condition \eqref{E-K1} on the gauge parameters $\kappa_0$ and $\kappa_1$  has been taken into account, and 
	\be
	\tilde b_{\mu\nu}\equiv b_{\mu\nu}-\frac{1}{3}\eta_{\mu\nu}b\label{def_tilde_b}
	\ee
is the traceless part of the $b_{\mu\nu}(x)$ field. From the above on-shell EoM we derive
	\begin{align}
	\eta_{\alpha\beta}\frac{\delta S}{\delta \hat a_{\alpha\beta}}&=-2ip_\mu \hat b^\mu-(k_0+3k_1)p^2\hat d=0\label{E-etaA}\\
	\frac{p_\alpha p_\beta}{p^2}\frac{\delta S}{\delta \hat a_{\alpha\beta}}&=(k_0+k_1)p^2\hat d=0\label{E-ppA}\\
	\frac{p_\alpha p_\beta}{p^2}\frac{\delta S}{\delta \hat{b}_{\alpha\beta}}&=\frac{2}{3}\kappa_0p_\mu \hat d^\mu=0\label{E-ppH}\\
	p^\alpha\frac{\delta S}{\delta \hat b^\alpha}&=-p^\mu p^\nu \hat a_{\mu\nu}+p^2\hat a=0\label{E-pV}\\
	p^\alpha\frac{\delta S}{\delta \hat d^\alpha}&=\kappa_0 p^\mu p^\nu \hat{b}_{\mu\nu}-\frac{1}{3}\kappa_0 p^2 \hat{b}=\kappa_0 p^\mu p^\nu\hat{\tilde{b}}_{\mu\nu}=0\ .\label{E-pB}
\end{align}
Using the conditions from the poles \eqref{E-poles}, according to which it must be $\kappa_0\neq0$ and $k_0\neq-k_1$, and the EoM   \eqref{E-etaA}, \eqref{E-ppA} and \eqref{E-ppH}, we get
	\begin{align}
	p^2\hat d&=0\label{E-p2b=0}\\
	p_\mu \hat d^\mu&=0\label{E-pb=0}\\
	p_\mu \hat b^\mu&=0\ .\label{E-pv=0}
	\end{align}
Moreover, from \eqref{E-eomPd} and \eqref{E-pV} we find
	\begin{align}
	p^\mu p^\nu \hat a_{\mu\nu}&=0\label{E-ppa=0}\\
	p^2\hat a&=0\ .\label{E-pTra=0}
	\end{align}
Considering \eqref{E-eomPa-gf}, we have
		\begin{align}
  \epsilon_{\alpha\lambda\rho}\frac{p^\lambda p_\beta}{p^2}\fdv{S}{\hat a_{\alpha\beta}} &=
  -\frac{i}{2}\left(
p^\mu \hat{\tilde{b}}_{\mu\rho}+\epsilon_{\rho\mu\nu}p^\mu \hat{b}^\nu
\right)=0\ ,\label{E-epsEomA}
		\end{align}
where we used \eqref{def_tilde_b} and the EoM \eqref{E-pB} with $\kappa_0\neq 0$. Comparing \eqref{E-epsEomA} with \eqref{E-eomPBvec} and considering the pole condition $\kappa_0\neq\kappa_2$, we get
\begin{align}
    p^\mu\hat{\tilde{b}}_{\mu\sigma}&=0 \label{pb=0dof}\\
    \epsilon_{\mu\nu\rho}p^\nu \hat b^\rho&=0\,.\label{epspb=0}
\end{align}
We observe that Eq. \eqref{pb=0dof} represents three equations on a 5-components field, thus the DoF contained in the traceless symmetric tensor $ b_{\mu\nu}(x)$ are \mbox{$5-3=2$}. Furthermore, \eqref{epspb=0} can be solved
	\be
	\hat b^\rho=p^\rho\hat \varphi\ ,
	\ee
where $\varphi(x)$, due to \eqref{E-pv=0}, is a scalar harmonic function, indicating that the vector field $\hat b^\mu(x)$ only contributes with one DoF. Finally, from \eqref{E-eomPv-gf} we have
	\be
	p^2\frac{\delta S}{\delta \hat b^\alpha}=- p^\mu p^2 \hat a_{\mu\alpha}+ \kappa_2\epsilon_{\alpha\mu\nu}p^2p^\mu \hat d^\nu=0\ ,
	\label{4.21}\ee
where we used \eqref{E-pTra=0}, and
	\be\label{E-epsEomH}
		\epsilon_{\beta\rho\sigma}\frac{p_\alpha p^\rho}{p^2}\frac{\delta S}{\delta \hat{b}_{\alpha\beta}}=
\epsilon_{\beta\rho\sigma} p^\rho
\left(\epsilon^{\beta\mu\nu}\tfrac{p_\mu p^\alpha}{p^2}\hat a_{\nu\alpha}+\kappa_0\hat d^\beta\right)
		=-p^\alpha \hat a_{\alpha\sigma}+\kappa_0\epsilon_{\sigma\mu\nu}p^\mu \hat d^\nu\ .
	\ee
Comparing \eqref{4.21} and \eqref{E-epsEomH},
using \eqref{E-pb=0}, \eqref{E-ppa=0} and the pole condition $\kappa_0\neq\kappa_2$ \eqref{E-poles}, we have	\begin{align}
	p^\alpha \hat a_{\alpha\sigma}&=0\label{4.23}\\
	\epsilon_{\alpha\mu\nu}p^\mu \hat d^\nu &=0\ .
	\end{align}
Eq.\eqref{4.23} represents three constraints on the 6-components of the symmetric tensor field $a_{\mu\nu}(x)$, which thus contribute with three DoF. We thus have three DoF from $ a_{\mu\nu}(x)$, two from $ b_{\mu\nu}(x)$ and one from $b^\mu(x)$, for a total of six DoF for the whole theory described by the gauge fixed action $S$ \eqref{def gauge fixed action}.

\section{Currents and fractons}\label{sec-currents}

To introduce matter in our theory, we generalize here at higher rank what is done, for instance, in Chern-Simons theory, where, following \cite{Dunne:1998qy,Tong:2016kpv}, matter current  is introduced by adding a term in the action
	\be
	S_{tot}=S_{CS}-\int d^3x A_\mu J^\mu\ ,
\label{CSmatter}	\ee
so that
	\be\label{J=eomCS}
	J^\mu=\frac{\delta S_{CS}}{\delta A_\mu}\ ,
	\ee
which encodes the matter response to electric and magnetic fields, since the 3D electric and magnetic fields are defined as \cite{Dunne:1998qy,Tong:2016kpv}
\begin{eqnarray}
\frac{\delta S_{CS}}{\delta A_a} &\propto& \epsilon^{0ab}E_b \label{tongdefE} \\
\frac{\delta S_{CS}}{\delta A_0} &\propto& B \label{tongdefB}\ .
\end{eqnarray}
For instance, the time component of \eqref{J=eomCS} relates the magnetic flux to the electric  charge density $J^0(x)$. Moreover, our theory depends on two fields, $a_{\mu\nu}(x)$ and $\tilde B_{\mu\nu}(x)$, hence we have two currents $J^{\mu\nu}(x)$ and $\tilde K^{\mu\nu}(x)$. This is in complete analogy with the BF description of 3D topological insulators (see for instance Eq. (14) of \cite{Cho:2010rk}). Finally, $\tilde K^{\mu\nu}(x)$ is traceless because it couples to $\tilde B_{\mu\nu}(x)$ \eqref{Btraceless}, which is traceless. Hence, the total action of our theory which includes matter is
	\be
	S_{tot}\equiv S_{BF} + S_J + S_K \,,\label{S_tot}
	\ee
where $S_{BF}$ is given by \eqref{S_BF} and
	\begin{align}
	S_J &\equiv -\int d^3 x\, J^{\mu\nu}a_{\mu\nu}\label{S_J}\\
	S_K &\equiv -\int d^3 x\, \tilde{K}^{\mu\nu}\tilde{B}_{\mu\nu}\label{S_k}\,,
	\end{align}
with $J^{\mu\nu}(x)$ a rank-2 symmetric tensor current, $\tilde{K}^{\mu\nu}(x)$ a rank-2 traceless tensor current and
$\tilde{B}_{\mu\nu}(x)$ is the traceless part of ${B}_{\mu\nu}(x)$ \eqref{Btraceless}.
The on-shell  EoM of the total action $S_{tot}$ \eqref{S_tot} are
	\begin{align}
	J^{\alpha\beta}&= \fdv{S_{BF}}{a_{\alpha\beta}}=\frac{1}{2}\left(
	\epsilon^{\mu\nu\alpha}\partial_\mu\invmixt{\tilde{B}}{\nu}{\beta}+\epsilon^{\mu\nu\beta}\partial_\mu\invmixt{\tilde{B}}{\nu}{\alpha}
	\right)\label{tensor current J}\\
	\tilde{K}^{\alpha\beta} &= \fdv{S_{BF}}{\tilde{B}_{\alpha\beta}}= \epsilon^{\mu\nu\alpha}\partial_\mu \invmixt{a}{\nu}{\beta} \label{tensor current k}\ .
	\end{align}
In vacuum, the 00-component of the on-shell EoM \eqref{tensor current J} of the symmetric field $a_{\mu\nu}(x)$  reads
	\be
	 \fdv{S_{BF}}{a_{00}}=\epsilon^{0mn}\partial_m\tilde B_n^{\ 0}=0 \,,
	\ee
which is solved by
	\be
	\tilde{B}_{j0} \equiv \partial_j \phi\ , \label{solB}
	\ee
where $\phi(x)$ is a local scalar function. This scalar field $\phi(x)$ plays the role of the one (typically called $A_0(x)$) introduced by hand in fracton theories \cite{ Pretko:2016lgv,Pretko:2016kxt}, and recovered as a vacuum solution in the covariant theories \cite{Bertolini:2022ijb,Bertolini:2024yur}. Thus it is at the heart of the fractonic interpretation. From now on, we will assume the solution \eqref{solB} to hold also when matter is introduced, in order to preserve the fractonic field content. As we shall see below, this is crucial in order to have a fractonic physical interpretation of our theory.  This assumption on the on-shell EoM \eqref{tensor current J} implies
	\be
	J^{00} =0 \label{J00=0}\ .
	\ee
From the on-shell EoM \eqref{tensor current J} we also get
	\be
	\partial_\alpha\partial_\beta J^{\alpha\beta} =0\ ,\label{ddJ=0}
	\ee
which, as a consequence of \eqref{J00=0},  can be rewritten as
	\be
	\partial_0 \rho + \partial_i \partial_j J^{ij}=0\label{fractonic continuity equation}\ ,
	\ee
where  we defined the charge density
	\begin{equation}
	\rho\equiv 2\partial_i J^{0i}\label{def_rho}\ .
	\end{equation}
We observe that \eqref{fractonic continuity equation} is a continuity equation for a scalar fractonic charge $\rho(x)$ since it implies both the conservation of the total charge
	\be
	\partial_0 \int  d \Sigma\, \rho = -\int d \Sigma\, \partial_i \partial_j J^{ij}=0 \,,
	\ee
where $ d \Sigma\equiv  d x_1  d x_2$, and of the total dipole momentum $D^k(t)$
	\be\label{dipole cons}
	\partial_0 D^k= \partial_0 \int  d \Sigma\, x^k \rho = -\int d \Sigma\,x^k \partial_i \partial_j J^{ij}= \int d \Sigma\, \partial_j J^{kj} =0\ ,  
	\ee
which encodes immobility, which is the fundamental property of fracton quasiparticles \cite{Pretko:2020cko}. Therefore we have a fully constrained fractonic charge and a fully mobile dipole excitation. Analogously to \eqref{tensor current J}, in vacuum the on-shell $00$-component of the EoM \eqref{tensor current k} is
		\be
		\frac{\delta S_{BF}}{\delta\tilde B_{00}}=\epsilon^{0mn}\partial_m a_n^{\ 0}=0\ ,
		\ee
which is solved by
		\be
		a_{n0}=\partial_n\psi\,,\label{sol B00}
		\ee
with $\psi(x)$ a local scalar function. This will be important for the fractonic interpretation of the theory because it will allow us to define, in analogy with the standard abelian BF \cite{Fradkin:2013sab,Cho:2010rk,Hansson:2004wca}, higher-rank electromagnetic fields, which are typical of fracton models \cite{ Pretko:2016lgv, Bertolini:2022ijb,Bertolini:2024yur}. In analogy to \eqref{solB}, from now on we will assume that the solution \eqref{sol B00} continues to be true when matter is introduced.  Thus, when using the solution \eqref{sol B00} in the $00$-component of the on-shell EoM \eqref{tensor current k}, it implies 
	\begin{equation}
	\tilde{K}^{00}=0\ ,\label{k00=0}
	\end{equation}
which, again, will play an important role in the physical interpretation of the theory. Moreover, from the on-shell EoM \eqref{tensor current k} we also get
	\begin{equation}
	\partial_\alpha\tilde{K}^{\alpha\beta}=0\ ,\label{dk=0}
	\end{equation}
whose components are explicitly given by
 \begin{itemize}
     \item $\beta=0$
     \begin{equation}
         \partial_i \tilde{K}^{i0}=0\,, \label{dki0=0}
     \end{equation}
     where \eqref{k00=0} has been used, and from which we observe that the vector quantity $\tilde{K}^{i0}(x)$ is solenoidal.
     \item $\beta=i$
     \begin{equation}
	\partial_0 \rho^{i} + \partial_j \tilde{K}^{ji}=0\label{cont-vect}\ ,
	\end{equation}
 which is a continuity equation for a vector charge density 
	\begin{equation}
	\rho^i\equiv \tilde{K}^{0i}\,,\label{def_rho_i}
	\end{equation}
with traceless current density $\tilde{K}^{ij}(x)$.
 \end{itemize}
Taking the divergence of the continuity equation \eqref{cont-vect} we also get a fractonic continuity equation as \eqref{fractonic continuity equation}
	\be
	\partial_0(\partial_i\rho^i)+\partial_i\partial_j\tilde K^{ij}=0\ ,
\label{fractoncontinuity}	\ee
where the role of the fractonic charge  is now played by $\partial_i\rho^i(x)$, and the symmetric part of the traceless tensor $\tilde K^{ij}(x)$ is a fractonic current. This implies that the vector charge $\rho^i(x)$ \eqref{def_rho_i} is a dipole-like quantity. Indeed, by definition of dipole momentum density $d^k(x)$, we have
	\be
	\int  d \Sigma\, d^k\equiv\int  d \Sigma\, x^k\partial_i\rho^i=-\int  d \Sigma\, \rho^k\ .
	\ee
From the continuity equation \eqref{cont-vect} we get
	\be
	\partial_0 \int  d \Sigma\, \rho^{i} = -\int  d \Sigma\, \partial_j \tilde{K}^{ji}=0\ ,
	\ee
which encodes the conservation of the vector charge. Using  \eqref{k00=0}, we also get
	\be
	\partial_0 \int  d \Sigma\, x_i \rho^i = -\int  d \Sigma\, x_i \partial_j \tilde{K}^{ji}= \int  d \Sigma\, \eta_{ij} \tilde{K}^{ji} = \int  d \Sigma\, \tilde{K}^{00} =0\ .\label{conservation law xiki0}
	\ee
  The physical meaning is the following: the vector charge density $\rho^i(x)$ \eqref{def_rho_i} is conserved and \eqref{conservation law xiki0} involves the trace of a quadrupole-like quantity $x^i \rho^j(x)$. 
In particular \eqref{conservation law xiki0} constrains the motion of the vector dipole-like charge to be transverse only \cite{Pretko:2016lgv,Pretko:2016kxt}.
  It is worth to note that from the continuity equation \eqref{cont-vect} we get
\begin{equation}
    \partial_0 \int  d \Sigma\, \epsilon_{0ij} x^i \rho^j = \int  d \Sigma\, \epsilon_{0ij}\tilde{K}^{ij} = \int  d \Sigma\, k_0 \ , \label{not_fractons}
\end{equation}
where we decomposed the current $\tilde{K}^{\alpha\beta}(x)$ into its symmetric and antisymmetric parts
\begin{equation}\label{Kdecomp}
    \tilde{K}^{\mu\nu} \equiv \tilde{k}^{\mu\nu}-\frac{1}{2}\epsilon^{\mu\nu\rho}k_\rho\ .
\end{equation}
Therefore, from \eqref{not_fractons} we observe that if $k_0(x)=0$ the theory displays an additional angular momentum-like conservation relation, and the dipole-like lineon becomes fractonic.
From \eqref{Kdecomp} we see that the condition $k_0(x)=0$ implies
\be
\tilde K^{ij}=\tilde k^{ij}\ ,
\ee
which is the symmetric component which intervenes in the fractonic continuity equation \eqref{fractoncontinuity}, and hence can be interpreted as a $fractonic$ dipole-like current. The spatially antisymmetric components of $\tilde B_{\mu\nu}(x)$, which are coupled to $\tilde K^{\mu\nu}(x)$ through $S_K$ \eqref{S_k}, are not physically relevant, and we may say that the condition $k_0=0$ implies a pure fractonic behaviour: both conserved charges of the theory, $\rho(x)$ \eqref{def_rho} and $\rho^i(x)$ \eqref{def_rho_i}, are associated to fractonic quasiparticles, and only dipolar bound states of $\rho(x)$ can move. On the other hand, when $k_0\neq0$ the angular momentum-like quantity \eqref{not_fractons} is not conserved, which allows the vector charge $\rho^i(x)$ \eqref{def_rho_i} to have a lineon-like behaviour.\\
Now, taking into account the solutions \eqref{solB} and \eqref{sol B00}, the non-trivial components of the on-shell EoM \eqref{tensor current k} read
	\begin{itemize}
	\item $\alpha=0$, $\beta=i$
		\be
		\rho^{i} =\frac{\delta S_{BF}}{\delta\tilde B_{0i}}= \epsilon^{0jk}\partial_j \invmixt{a}{k}{i}\,.\label{EoM_rho_i}
		\ee
	\item $\alpha=i$, $\beta=j$ 
		\be
		\tilde{K}^{ij} =\frac{\delta S_{BF}}{\delta\tilde B_{ij}}= \epsilon^{0ki}\partial_0\invmixt{a}{k}{j}+\epsilon^{0ik}\partial_k\invmixt{a}{0}{j}\,.\label{EoM_K_ij}
		\ee
	\end{itemize}
We recall that in ordinary 3D BF theory \eqref{bf}, one has
	\begin{align}
	\fdv{S_{BF}^{(ord)}}{B_i} &\propto \epsilon^{0ij} \mathcal E_j\\
	\fdv{S_{BF}^{(ord)}}{B_0} &\propto \mathcal B \,,
	\end{align}
with $\mathcal E_i(x)$ and $ \mathcal B(x)$ being the planar electric and magnetic fields \cite{Fradkin:2013sab,Cho:2010rk,Hansson:2004wca}. Analogously, here we define
	\begin{align}
	\fdv{S_{BF}}{\tilde{B}_{ij}}\bigg|_{\mbox{\tiny\eqref{sol B00}}} &\equiv \frac{1}{2}\epsilon^{0ik} \invmixt{\mathcal E}{k}{j}\label{def_Eij}\\
	\fdv{S_{BF}}{\tilde{B}_{0i}} &\equiv \frac{1}{2} \mathcal B^i\,,\label{def_Bi}
	\end{align}
where $ \mathcal E_{ij}(x)$ and $\mathcal B_i(x)$ are generalized electric and magnetic fields. These, by comparing with \eqref{tensor current J} and \eqref{tensor current k} in vacuum, can be written in terms of the fracton field strength \eqref{def_fracton_field_strength} as
	\begin{align}
	\mathcal E_{ij} &\equiv F_{ij0} \label{Eij_from_F}\\
	\mathcal B_i &\equiv\frac{2}{3} \epsilon^{0jk} F_{ijk}\,.\label{Bi_from_F}
	\end{align}
Notice that, due to \eqref{Eij_from_F} and to the solution \eqref{sol B00}, the generalized electric field $\mathcal E_{ij}(x)$ is symmetric as in fractonic theories
 \cite{Prem:2017kxc,Bertolini:2022ijb,Pretko:2017xar}.
Furthermore, using the definitions \eqref{def_Eij} and \eqref{def_Bi}, the EoM \eqref{EoM_rho_i} and \eqref{EoM_K_ij} can be rewritten, respectively, as
	\begin{equation}
	\rho^i = \frac{1}{2}\mathcal B^i \label{vector_charge_flux_attachment}
	\end{equation}
and
	\begin{equation}
	\tilde{K}^{ij}= \tilde{\sigma}^{ijkl}\mathcal  E_{kl}\ ,\label{generalized_Hall_current}
	\end{equation}
with 
	\begin{equation}
	\tilde\sigma^{ijkl}=\tilde\sigma^{ijlk} \equiv \frac{1}{4}\left(
	\epsilon^{0ik}\eta^{jl} + \epsilon^{0il}\eta^{jk}
	\right)\ ,
	\end{equation}
which is traceless on the first two indices. Notice also that \eqref{vector_charge_flux_attachment} and \eqref{generalized_Hall_current} are, respectively, a generalization of the magnetic flux attachment relation
\begin{equation}
    \rho \propto \mathcal{B}
\end{equation}
and of the Hall current
\begin{equation}
    J^i \propto \epsilon^{0ij} \mathcal E_j \,,\label{Hall_current_BF}
\end{equation}
which characterize both the ordinary Chern-Simons \eqref{cs} and BF \eqref{bf} actions when coupled to matter \cite{Fradkin:2013sab,Cho:2010rk,Hansson:2004wca}.
Notice that the 0-component of the EoM of the antisymmetric part, represented by the vector $b^\mu(x)$, coupled to its current $k_\mu$ \eqref{Kdecomp} reads
	\be\label{k0=E}
	k_0=\frac{\delta S_{BF}}{\delta b^0}=-\partial^\mu a_{\mu0}+\partial_0a\\
		=-\frac{1}{2}F^m_{\ m0}=-\frac{1}{2}\mathcal E^m_{\ m}\ ,
	\ee
where we used the definition of the electric tensor field \eqref{Eij_from_F}. A physical interpretation of the lineon-to-fracton transformation thus emerges and is the following: the general theory \eqref{S_tot} features a scalar fracton and a vectorial lineon whose motion is associated to electromagnetic-like fields $\mathcal E_{ij}(x)$ \eqref{Eij_from_F} and $\mathcal B^i(x)$ \eqref{Bi_from_F}  through the Hall-like relations \eqref{generalized_Hall_current} and \eqref{vector_charge_flux_attachment} respectively. A transition happens when the trace of the electric tensor $\mathcal E_{ij}(x)$ is turned off ($i.e.$ $k_0=0$ on-shell in \eqref{k0=E}), for which the system acquires an angular-momentum-like conservation and the lineon becomes a fracton. In other words the trace of the electric tensor $\mathcal E^m_{\ m}$ is related to the breaking of angular momentum. This transition can be seen as stepping from the so called ``vector charge theory of fractons'' to the ``traceless vector charge theory of fractons'' \cite{Pretko:2016lgv}.
To conclude, a comment on the connection  between this theory and the existing literature is in order. Using the vacuum solutions of the on-shell EoM \eqref{tensor current J} and \eqref{tensor current k} for $a_{00}(x)$ and $\tilde{B}_{\alpha0}(x)$, thus implying $\tilde K_{i0}(x)=0$ in addition to \eqref{J00=0} and \eqref{k00=0}, these  fields can be integrated out from the partition function associated to the total action $S_{tot}$ \eqref{S_tot}, which leads to the effective action
	\be
	S_{eff} 
	= \int d^3x\,\big(\psi\epsilon^{0ij}\partial_{j}\partial_k\invmixt{\tilde{B}}{i}{k}  +  \invmixt{\tilde{B}}{0}{k}\epsilon^{0ij}\partial_{i} a_{jk}-a_{ik}\epsilon^{0ij}\partial_0\invmixt{\tilde{B}}{j}{k}
	-\tilde{B}_{0i}\rho^i
	-\tilde{B}_{ij}\tilde{K}^{ij}+
	\psi\rho-a_{ij} J^{ij}
	\big)\,,\label{action_R2TC}
	\ee
which can be mapped into Eq. (3.32) of \cite{Han:2024nvu}. Thus in this case our theory describes the low-energy limit of the Rank-2 Toric Code (R2TC) in two spatial dimensions \cite{Oh:2021gee,Oh:2022klh}, which is an exactly solvable quantum lattice model whose quasiparticle excitations have restricted mobility and exhibit unusual braiding statistics \cite{Oh:2021gee, Pace:2022wgl, Oh:2023bnk}. Explicitly the mapping is the following
	\bea
	\mbox{\bf R2TC}&&\mbox{\bf Covariant BF-like}\nonumber\\
	(  E^x_t,\,  E^y_t) &\equiv& (-\tilde{B}_{02},\,\tilde{B}_{01})\label{def Ext Eyt}\\
	(  E_{xx},  E_{yy},   E_{xy})&\equiv& (-\tilde{B}_{12},\,\tilde{B}_{21},\,\tilde{B}_{11}-\tilde{B}_{22}) \label{def Eij}\\	(A_t,A_{xx},A_{yy},A_{xy})&\equiv& (\psi,\,a_{22},\, a_{11},\, -a_{12}) \label{def_Aij Han}\\
	( \tilde{J}_t,\,  {J}_{xx},\,  {J}_{xy},\,  {J}_{yy}) &\equiv& -(\rho,\, J_{11}, \, 2J_{12},\, J_{22})\label{def Jtilde}\\
	( {K}_{xx},\, {K}_{xy},\,  {K}_{yy}) &\equiv& (-\tilde{K}_{12},\,\tilde{K}_{11},\, \tilde{K}_{21})\label{def Ktilde}\\
	(\rho^x,\, \rho^y) &\equiv& (-\rho_2,\, \rho_1)\ ,\label{def rhox rhoy}
	\eea
from which we observe that the components of the vector charge density $\rho_i(x)$ correspond to the two magnetic excitations which characterize the R2TC. Moreover, we stress that this relation between the covariant BF-like theory and the R2TC is a higher-rank generalization of what is proved in \cite{Slagle:2017wrc}, where an equivalence between the ordinary 3D BF theory and the Kitaev's Toric Code \cite{Kitaev:1997wr} has been demonstrated.
Finally, since the R2TC action is equivalent to the dipolar BF action studied in \cite{Ebisu:2023idd}, then our theory is also equivalent to the  foliated BF theory with global and dipole symmetry of \cite{Ebisu:2023idd} (see in particular Eq. (3.7)).

\section{Symmetric tensor fields}\label{Sec-symm}

As a particular case, we now consider the theory 
where also the tensor field $B_{\mu\nu}(x)$ appearing in the action $S_{BF}$ \eqref{Sinv} is symmetric. To avoid confusion with the generic case, we call this latter field $\Phi_{\mu\nu}(x)$. Therefore,  both the tensor fields $a_{\mu\nu}(x)$ and $\Phi_{\mu\nu}(x)$ are symmetric
\bea
a_{\mu\nu} &=& a_{\nu\mu}\\
\Phi_{\mu\nu} &=&\Phi_{\nu\mu}\ ,
\label{Bsimm}\eea
and whose transformations are
	\bea
	\delta'_1 a_{\mu\nu} = \partial_{\mu}\partial_{\nu}\Lambda\quad &;& \quad 
	\delta'_1\Phi_{\mu\nu} = 0  \label{gauge transformation 1 symm}\\
	\delta'_2 a_{\mu\nu} = 0 \quad &;& \quad 
	\delta'_2 \Phi_{\mu\nu} =\partial_\mu \partial_\nu \xi\ , \label{gauge transformation 2 symm}
	\eea
where $\Lambda(x)$ and $\xi(x)$ are two local scalar gauge parameters. As done for the action \eqref{Sinv}, requiring vanishing ${\cal P}$-charge \eqref{Pcharge}, we get the most general invariant action
 \begin{align}
     S^{(s)}_{BF} &= \int d^3 x\, \epsilon^{\mu\nu\rho}\Phi_{\mu\sigma}\partial_\nu a_\rho^{\ \sigma}\ ,\label{S_BF-symm}
 \end{align}
 which satisfies
 \be
 \delta'_1S^{(s)}_{BF}  = \delta'_1S^{(s)}_{BF}  = {\cal P}S^{(s)}_{BF}  = 0\ .
 \ee
 The two longitudinal diffeomorphisms transformations \eqref{gauge transformation 1 symm} and \eqref{gauge transformation 2 symm} require two scalar gauge fixing conditions of the type \eqref{scalar gauge fixing}, and the gauge fixing procedure  straightforwardly follows what we have already done in Section 2.
It is worth to remark that, as in the standard abelian case \cite{Birmingham:1991ty}, the BF-like action \eqref{S_BF-symm} can be cast into the sum of two Chern-Simons-like actions, which is 
 a rank-2 generalization of what happens in the ordinary abelian 3D BF theory, where the action \eqref{bf} results from the combination of two Chern–Simons actions \eqref{cs} with opposite chiralities \cite{Birmingham:1991ty}.
In fact, by means of the linear transformation
  \begin{align}
     a^\pm_{\mu\nu} \equiv a_{\mu\nu} \pm \Phi_{\mu\nu} \,,
 \end{align}
the action \eqref{S_BF-symm} becomes
\begin{equation}
    S^{(s)}_{BF} = \frac{1}{4}\left(
    S^+_{CS}-S^-_{CS}
    \right)\,, \label{BFdoubleCS}
\end{equation}
where
\begin{align}
    S^\pm_{CS} \equiv \int d^3 x\, \epsilon^{\mu\nu\rho}a^{\pm\sigma}_{\,\mu}\partial_\nu a^\pm_{\rho\sigma}\,.\label{CSlike}
\end{align}
The single Chern-Simons-like action was recently studied in \cite{Bertolini:2024yur}.
Therefore, due to the relation \eqref{BFdoubleCS}, the Hall-like interpretation of the fractonic Chern-Simons-like action \eqref{CSlike}, discussed in \cite{Bertolini:2024yur}, holds for the action \eqref{S_BF-symm} as well. 
When matter is introduced, the total action reads
\be
	S_{tot}^{(s)}\equiv S_{BF}^{(s)} + S_J^{(s)} + S_K^{(s)}\,, \label{S_tot-symm}
	\ee
where $S_{BF}^{(s)}$ is given by \eqref{S_BF-symm} and
	\begin{align}
	S_J^{(s)} &\equiv -\int d^3 x\, \tilde{J}^{\mu\nu}\tilde{a}_{\mu\nu}\label{S_J-symm}\\
	S_K^{(s)} &\equiv -\int d^3 x\, \tilde{k}^{\mu\nu}\tilde{\Phi}_{\mu\nu}\label{S_k-symm}\,,
	\end{align}
where $\tilde{a}_{\mu\nu}(x)$ and $\tilde{\Phi}_{\mu\nu}(x)$ are the traceless components of $a_{\mu\nu}(x)$ and
$\Phi_{\mu\nu}(x)$ while $\tilde{J}^{\mu\nu}(x)$ and $\tilde{k}^{\mu\nu}(x)$ are rank-2 symmetric traceless tensor currents. The corresponding on-shell EoM are
\begin{align}
    \tilde{J}^{\alpha\beta}&= \fdv{S_{BF}^{(s)}}{\tilde{a}_{\alpha\beta}}=\frac{1}{2}\left(
\epsilon^{\mu\nu\alpha}\partial_\mu\invmixt{\tilde{\Phi}}{\nu}{\beta}+\epsilon^{\mu\nu\beta}\partial_\mu\invmixt{\tilde{\Phi}}{\nu}{\alpha}
    \right)\label{tensor current J-symm}\\
    \tilde{k}^{\alpha\beta} &= \fdv{S_{BF}^{(s)}}{\tilde{\Phi}_{\alpha\beta}}= \frac{1}{2}\left(
\epsilon^{\mu\nu\alpha}\partial_\mu\invmixt{\tilde{a}}{\nu}{\beta}+\epsilon^{\mu\nu\beta}\partial_\mu\invmixt{\tilde{a}}{\nu}{\alpha}
    \right) \,, \label{tensor current k-symm}
\end{align}
which we observe to be a rank-2 generalization of the ones derived for the description of topological insulators \cite{Cho:2010rk}. From \eqref{tensor current J-symm} and \eqref{tensor current k-symm} and assuming that the vacuum solutions of the EoM of $\tilde{a}_{00}(x)$ and $\tilde{\Phi}_{00}(x)$, given by 
\begin{align}
    \tilde{a}_{0j} &= \partial_j \psi \label{solPhi00-symm}\\
    \tilde{\Phi}_{0j} &= \partial_j \phi \label{sola00-symm}\,,
\end{align}
continue to be true also when matter is introduced, we can derive the two continuity equations
\begin{align}
    \partial_0 \rho_{\mbox{\tiny J}}  + \partial_i \partial_j \tilde{J}^{ij}&=0\label{fractonJ_cont_eq-symm}\\
    \partial_0 \rho_{\mbox{\tiny K}}  + \partial_i \partial_j \tilde{k}^{ij}&=0\,,\label{fractonK_cont_eq-symm}
\end{align}
where we defined the two scalar charge densities
\begin{align}
    \rho_{\mbox{\tiny J}} &= 2\partial_i \tilde{J}^{i0}\label{def_rhoJ}\\
    \rho_{\mbox{\tiny K}} &= 2\partial_i \tilde{k}^{i0}\,. \label{def_rhoK}
\end{align}
These charge densities are fractonic since \eqref{fractonJ_cont_eq-symm} and \eqref{fractonK_cont_eq-symm} imply both the conservation of total charges and of the total dipole momenta. Moreover, from the continuity equations \eqref{fractonJ_cont_eq-symm} and \eqref{fractonK_cont_eq-symm}, the traces of the quadrupole momenta are conserved 
\begin{align}
    \partial_0 \int  d \Sigma\, \eta^{ij}x_i x_j \rho_{\mbox{\tiny J}} &= -\int  d \Sigma\, \eta^{ij}x_i x_j \partial_k \partial_l \tilde{J}^{kl} \propto \int  d \Sigma\, \eta_{kl} \tilde{J}^{kl}= \int  d \Sigma\,  \tilde{J}^{00} =0\label{cons_trQuadrupoleJ}\\
    \partial_0 \int  d \Sigma\, \eta^{ij}x_i x_j \rho_{\mbox{\tiny K}} &= -\int  d \Sigma\, \eta^{ij}x_i x_j \partial_k \partial_l \tilde{k}^{kl} \propto\int  d \Sigma\, \eta_{kl} \tilde{k}^{kl}=\int  d \Sigma\, \tilde{k}^{00} =0\,,\label{cons_trQuadrupoleK}
\end{align}
where the last steps follow from the solutions \eqref{solPhi00-symm} and \eqref{sola00-symm}, which imply that the $00$-components of the traceless tensor currents $\tilde{k}^{\alpha\beta}(x)$ and $\tilde{J}^{\alpha\beta}(x)$ vanish. Hence both dipoles are constrained to move in straight lines which are perpendicular to their dipole momenta \cite{Doshi}, \textit{i.e.} they behave as lineons. Working on-shell on the vacuum solution \eqref{solPhi00-symm} of the EoM of $\tilde{\Phi}_{00}(x)$, in analogy with the standard abelian BF theory in 3D \cite{Fradkin:2013sab,Cho:2010rk,Hansson:2004wca}, we can rewrite the on-shell EoM \eqref{tensor current k-symm}  explicitly in terms of generalized electric and magnetic fields 
\begin{align}
    \fdv{S^{(s)}_{BF}}{\tilde{\Phi}_{ij}}\bigg|_{ \mbox{\tiny\eqref{solPhi00-symm}}} &\equiv \frac{1}{4}\left(\epsilon^{0ik}\invmixt{\tilde{\mathcal E}}{k}{j}+\epsilon^{0jk}\invmixt{\tilde{\mathcal E}}{k}{i}
    \right)\label{Etraceless}\\
    \fdv{S^{(s)}_{BF}}{\tilde{\Phi}_{0i}} &\equiv \frac{1}{2}\tilde{\mathcal{B}}^i
\end{align}
as
\begin{align}
    \rho_{\mbox{\tiny K}} &= \partial_i \tilde{\mathcal{B}}^i\label{Gauss-symm}\\
    \tilde{K}^{ij} &= \tilde{\sigma}^{ijkl}_{\mbox{\tiny (s)}}\tilde{\mathcal E}_{kl}\,,\label{Hall-symm}
\end{align}
with
\begin{equation}
    \tilde{\sigma}^{ijkl}_{\mbox{\tiny (s)}}= \tilde{\sigma}^{ijlk}_{\mbox{\tiny (s)}}=\tilde{\sigma}^{jikl}_{\mbox{\tiny (s)}}\equiv \frac{1}{8}\left(
    \epsilon^{0ik}\eta^{jl} + \epsilon^{0il}\eta^{jk} + \epsilon^{0jk}\eta^{il} + \epsilon^{0jl}\eta^{ik}
    \right)\,.\label{def_sigma-symm}
\end{equation}
Notice that \eqref{Gauss-symm} and \eqref{Hall-symm} represent, respectively, a magnetic Gauss law and a generalized Hall current for the type-K charges. To conclude, the completely symmetric case studied in this Section corresponds to two fractonic traceless scalar charge theories and it is a rank-2 generalization of the action proposed in \cite{Cho:2010rk},  where it has been proved that ordinary 3D BF theory is a good effective field theory for the description of  quantum spin Hall insulators in two spatial dimensions.
Recently in \cite{Lam:2024smz}, in a related context of dipole conserving theories, a bulk description of topological dipole insulators has been proposed  starting from an effective edge theory and by means of so-called coupled wire construction \cite{Meng:2019ket}. In this approach, they made the assumption that the dipole momentum is conserved {\it only} in one direction, say the $x_1$ direction. Related to this, is the fact that the gauge fields of the resulting bulk theory appearing in \cite{Lam:2024smz} do not transform under the covariant fracton gauge transformations \eqref{gauge transformation 1 symm} and \eqref{gauge transformation 2 symm}. Our theory generalizes the one studied in \cite{Lam:2024smz} since, from \eqref{fractonJ_cont_eq-symm} and \eqref{fractonK_cont_eq-symm}, we have conservation of both the $x_1$ and $x_2$ components of the total dipole momentum. This is a consequence of the covariance of our fracton gauge theory.
In terms of subdimensional quasiparticles, there are no fractons in \cite{Lam:2024smz} but only scalar lineons, defined by the conservation of the dipolar momentum component transverse to their propagation direction. Unlike \cite{Lam:2024smz}, as we showed in this Section, our theory is characterized by two types of fractons and lineons. Importantly, the latter are a consequence of the tracelessness of our theory and are given by the dipoles, which are constrained to move in the direction orthogonal to their dipole momenta. A detailed study of the resulting edge theory in our case is an interesting direction to be inspected. 

\section{Conclusions}\label{sec-conclusion}

Fracton models in 3D emerge in many condensed matter contexts where mobility constraints or multipole conservations are present. Examples are, for instance, the elasticity duality for topological defects and hydrodynamics, or cases which display subsystem symmetries and dipolar behaviours \cite{Prem:2017kxc, Gromov:2022cxa}. In most of these situations 3D non-covariant higher-rank models come into play \cite{Huang:2023zhp,Pretko:2017kvd}. Following the construction of the higher-rank covariant Chern-Simons model for fractons \cite{Bertolini:2024yur},
in this paper we investigated the possibility of covariant 3D fractonic BF models. \\

We thus started by considering a theory defined by two rank-2 tensor fields, one of which -$a_{\mu\nu}(x)$- transforms under the covariant fracton symmetry \eqref{gauge transformation 1}, while the second -$B_{\mu\nu}(x)$- obeys the more general electromagnetic-like transformation with a vector gauge parameter \eqref{gauge transformation 2}. The most general invariant action  generated by these symmetries  and involving both fields, has indeed a BF-like form \eqref{S_BF}. Differently from the standard 3D model $S_{BF}^{(ord)}$ \eqref{bf}, the invariant action $S_{BF}$ \eqref{S_BF} is not topological, due to a linear dependence on the metric. This is similar to what happens in the covariant higher-rank Chern-Simons-like model \cite{Bertolini:2024yur}, from which, however, it differs by having an on-shell vanishing energy-momentum tensor, making the BF-like model \eqref{S_BF} ``quasi-topological''. \\
Additionally, the action $S_{BF}$ \eqref{S_BF} does not depend on the trace of the non-symmetric tensor field $B_{\mu\nu}(x)$, and indeed in order to have propagators, the gauge fixing must not depend on the trace of $B_{\mu\nu}(x)$. This has direct consequences on the number of DoF. In fact three DoF come from the symmetric field $a_{\mu\nu}(x)$ and three from the non-symmetric one $B_{\mu\nu}(x)$. Comparing this with the fully symmetric and fully traceless case studied in Section \ref{Sec-symm},
there is a difference of two DoF, which is a consequence of the presence of the non-symmetric field instead of a purely symmetric one, affecting the physical content of the two theories. \\

A first hint towards a physical interpretation of the model described by the action $S_{BF}$ \eqref{S_BF} appears when generalized electric and magnetic fields $\mathcal E^{ij}(x)$ \eqref{Eij_from_F} and $\mathcal B^i(x)$ \eqref{Bi_from_F} are defined, in analogy to ordinary electromagnetism \cite{Pretko:2016lgv, Bertolini:2022ijb}, in terms of the invariant fracton field strength $F_{\mu\nu\rho}(x)$ \eqref{def_fracton_field_strength}. 
The existence of higher-rank electromagnetic-like fields is a first sign that the theory is fractonic, since it exhibits a form of generalized, higher-rank electromagnetism  \cite{Pretko:2016lgv}. Notice that, as a consequence of having considered a non-symmetric field, here the electric-like field $\mathcal E^{ij}(x)$ \eqref{Eij_from_F} is not traceless, in contrast to what happens in the theory described by $S^{(s)}_{BF}$ \eqref{S_BF-symm}, and in the Chern-Simons-like model \cite{Bertolini:2024yur}, for which the electric field $\tilde{\mathcal E}^{ij}(x)$ \eqref{Etraceless} is traceless.\\
This fact, which might appear rather formal, has indeed a physical consequence, since it is relevant  in determining the mobility of the fracton quasiparticles, and in fact the two models display a different quasiparticle content and different conservations when fractonic matter is taken into account. Indeed the full physical content of the model arises, and its fractonic behaviour emerges, only when matter is introduced. When matter is coupled to the pure gauge theory  $S_{BF}$ \eqref{S_BF} through $S_{tot}$ \eqref{S_tot}, the continuity equations \eqref{fractonic continuity equation} and \eqref{cont-vect} that constrain the motion of the quasiparticles through conservation relations, are found, and the following fundamental conserved charges can be identified 
	\bi
	\item a scalar charge $\rho(x)$ \eqref{def_rho}, which is of the \textbf{fractonic} type, as a consequence of the dipole conservation \eqref{dipole cons}, and whose dipolar bound states are free to move;
	\item a vector dipole-like charge $\rho^i(x)$ \eqref{def_rho_i} which has two possible behaviours depending on the 0-component of the current $k_\mu(x)$ \eqref{Kdecomp}. In particular, we can identify the vector charge as associated to a {\bf lineon}-like behaviour thanks to the conservation of a quadrupole momentum-like component \eqref{conservation law xiki0}, which constrains the quasiparticle to move on a line. However, whenever $k_0(x)=0$, the additional angular momentum-like conservation \eqref{not_fractons} appears, which further constrains the dipole-like charge $\rho^i(x)$ to be immobile, $i.e.$ purely {\bf fractonic}.
	\ei
This distinguishes the case described by the action $S_{BF}$ \eqref{S_BF} from the one described by $S_{BF}^{(s)}$ \eqref{S_BF-symm}, for which both charges $\rho_{\mbox{\tiny J,K}}(x)$ \eqref{def_rhoJ} and \eqref{def_rhoK} are scalar and fractonic. The associated dipole momenta $x^i\rho_{\mbox{\tiny J,K}}(x)$ move on a line as a consequence of the quadrupole-like conservations \eqref{cons_trQuadrupoleJ} and \eqref{cons_trQuadrupoleK}, whose existence is related to the fact that both fields of the theory are traceless. Additionally, dipole-like flux attachment relations \eqref{vector_charge_flux_attachment} and \eqref{Gauss-symm}, and Hall-like conductivities \eqref{generalized_Hall_current} and \eqref{Hall-symm} are observed for both models $S_{BF}$ \eqref{S_BF} and $S_{BF}^{(s)}$ \eqref{S_BF-symm}. \\

Furthermore, in the case of $S_{BF}$ \eqref{S_BF}, together with the continuity equations \eqref{fractonic continuity equation} and \eqref{cont-vect}, a solenoidal condition \eqref{dki0=0} is recovered for the $i0$-component of the current $\tilde K_{\mu\nu}(x)$. When this condition is trivially satisfied, $i.e.$ when $\tilde K^{i0}(x)=0$, the higher-rank BF-like model \eqref{S_BF} can be cast into the action $S_{eff}$ \eqref{action_R2TC} which is the effective field theory of the R2TC \cite{Oh:2022klh, Han:2024nvu} through the mapping \eqref{def Ext Eyt}-\eqref{def rhox rhoy}. 
In this context the dipole-like vector charge $\rho^i(x)$ \eqref{def_rho_i},  can be interpreted
as the magnetic excitations of the R2TC as a consequence of the mapping \eqref{def rhox rhoy}. 
Additionally, from the flux-attachment relation \eqref{vector_charge_flux_attachment} it is also possible to relate our magnetic field $\mathcal{B}^a(x)$ \eqref{Bi_from_F} 
to the one of the rank-2 U(1) lattice gauge theory, 
connected to the R2TC \cite{Oh:2021gee,Oh:2022klh} through a ``Higgsing procedure'' \textcolor{black}{on the lattice} \cite{Bulmash:2018lid}.\\

Topological Chern-Simons and BF models have a long history of important physical results when boundaries are introduced \cite{Cho:2010rk,Wen:1990se,Amoretti:2014iza,Maggiore:2018bxr,Maggiore:2017vjf,Bertolini:2021iku,Bertolini:2022sao}. The higher-rank similarities shared by the BF-like models \eqref{S_BF} and \eqref{S_BF-symm} thus suggest possible interesting perspectives for an analysis of boundary phenomena. In particular, the case of $S_{BF}^{(s)}$ \eqref{S_BF-symm} generalizes the recently proposed bulk description of topological dipole insulators \cite{Lam:2024smz} where, starting from an effective edge theory by means of so-called coupled wire construction \cite{Meng:2019ket},
a non-covariant BF-like model is recovered. A study of the corresponding boundary action of the covariant $S_{BF}^{(s)}$ \eqref{S_BF-symm} is then worthwhile. Moreover, it can also be interesting to investigate boundary effects for $S_{BF}$ \eqref{S_BF}, where non symmetric contribution might lead to non trivial physics.

\section*{Acknowledgements}

 M.C. acknowledges support from the project PRIN 2022 - PH852L(PE3)
TopoFlags  funded by the European community - Next Generation EU
within the programme "PNRR Missione 4 - Componente 2 - Investimento 1.1
Fondo per il Programma Nazionale di Ricerca e Progetti di Rilevante
Interesse Nazionale (PRIN)".

\section*{Author contribution statement}

All authors contributed equally to this work, including writing and revision of the manuscript.

\appendix

\section{Calculation of the propagators}\label{App:propagators}

The propagators appearing in the $\Delta_{AP}(p)$ matrix \eqref{Delta} can be written as
\allowdisplaybreaks
	\begin{align}
	\Delta^{_{(1)}}_{\alpha\beta,\rho\sigma}(p)&\equiv \langle \hat a_{\alpha\beta}(-p)\,\hat  a_{\rho\sigma}(p)\rangle\\
	&=\left(c_0A^{(0)}+c_1A^{(1)}+c_2A^{(2)}+c_3A^{(3)}+c_4A^{(4)}+c_5A^{(5)}+c_6A^{(6)}\right)_{\alpha\beta,\rho\sigma}\nonumber\\
	\Delta^{_{(2)}}_{\alpha\beta,\rho\sigma}(p)&\equiv \langle \hat a_{\alpha\beta}(-p)\, \hat{b}_{\rho\sigma}(p)\rangle\\
	&=\left(c_{7}A^{(0)}+c_{8}A^{(1)}+c_{9}A'^{(2)}+c_{10}B^{(2)}+c_{11}A^{(3)}+c_{12}A^{(4)}+c_{13}A^{(5)}+c_{14}A^{(6)}\right)_{\alpha\beta,\rho\sigma}\nonumber\\
	\Delta^{_{(3)}}_{\alpha\beta,\rho}(p)&\equiv \langle \hat a_{\alpha\beta}(-p)\, \hat b_{\rho}(p)\rangle\\
	&= ic_{15}\,\eta_{\alpha\beta}p_\rho+ic_{16}\,p_\alpha p_\beta p_\rho+c_{17}\,p^\lambda(\epsilon_{\alpha\lambda\rho}p_\beta+\epsilon_{\beta\lambda\rho}p_\alpha)+ic_{18}\left(\eta_{\alpha\rho}p_\beta+\eta_{\beta\rho}p_\alpha\right)\nonumber\\
	\Delta^{_{(4)}}_{\alpha\beta,\rho}(p)&\equiv \langle \hat a_{\alpha\beta}(-p)\, \hat d_{\rho}(p)\rangle\\
	&= ic_{19}\,\eta_{\alpha\beta}p_\rho+ic_{20}\,p_\alpha p_\beta p_\rho+c_{21}\,p^\lambda(\epsilon_{\alpha\lambda\rho}p_\beta+\epsilon_{\beta\lambda\rho}p_\alpha)+ic_{22}\left(\eta_{\alpha\rho}p_\beta+\eta_{\beta\rho}p_\alpha\right)\nonumber\\
	\Delta^{_{(5)}}_{\alpha\beta}(p)&\equiv \langle \hat a_{\alpha\beta}(-p)\, \hat d(p)\rangle\\
	&= c_{23}\,\eta_{\alpha\beta}+c_{24}\,p_\alpha p_\beta\nonumber\\
	\Delta^{_{(6)}}_{\alpha\beta,\rho\sigma}(p)&\equiv \langle \hat{b}_{\alpha\beta}(-p)\, \hat{b}_{\rho\sigma}(p)\rangle\\
	&=\left(c_{25}A^{(0)}+c_{26}A^{(1)}+c_{27}A^{(2)}+c_{28}A^{(3)}+c_{29}A^{(4)}+c_{30}A^{(5)}+c_{31}A^{(6)}\right)_{\alpha\beta,\rho\sigma}\nonumber\\
	\Delta^{_{(7)}}_{\alpha\beta,\rho}(p)&\equiv \langle \hat{b}_{\alpha\beta}(-p)\, \hat b_{\rho}(p)\rangle\\
	&= ic_{32}\,\eta_{\alpha\beta}p_\rho+ic_{33}\,p_\alpha p_\beta p_\rho+c_{34}\,p^\lambda(\epsilon_{\alpha\lambda\rho}p_\beta+\epsilon_{\beta\lambda\rho}p_\alpha)+ic_{35}\left(\eta_{\alpha\rho}p_\beta+\eta_{\beta\rho}p_\alpha\right)\nonumber\\
	\Delta^{_{(8)}}_{\alpha\beta,\rho}(p)&\equiv \langle \hat{b}_{\alpha\beta}(-p)\, \hat d_{\rho}(p)\rangle\\
	&= ic_{36}\,\eta_{\alpha\beta}p_\rho+ic_{37}\,p_\alpha p_\beta p_\rho+c_{38}\,p^\lambda(\epsilon_{\alpha\lambda\rho}p_\beta+\epsilon_{\beta\lambda\rho}p_\alpha)+ic_{39}\left(\eta_{\alpha\rho}p_\beta+\eta_{\beta\rho}p_\alpha\right)\nonumber\\
	\Delta^{_{(9)}}_{\alpha\beta}(p)& \equiv\langle \hat{b}_{\alpha\beta}(-p)\,\hat d(p)\rangle\\
	&= c_{40}\,\eta_{\alpha\beta}+c_{41}\,p_\alpha p_\beta\nonumber\\
	\Delta^{_{(10)}}_{\alpha,\rho}(p)&\equiv \langle \hat b_{\alpha}(-p)\,\hat b_{\rho}(p)\rangle\\
	&= c_{42}\,\eta_{\alpha\rho}+c_{43}\,p_\alpha p_\rho+ic_{44}\,p^\lambda\epsilon_{\alpha\lambda\rho}\nonumber\\
	\Delta^{_{(11)}}_{\alpha,\rho}(p)&\equiv  \langle \hat b_{\alpha}(-p)\, \hat d_{\rho}(p)\rangle\\
	&=c_{45}\,\eta_{\alpha\rho}+c_{46}\,p_\alpha p_\rho+ic_{47}\,p^\lambda\epsilon_{\alpha\lambda\rho}\nonumber\\
	\Delta^{_{(12)}}_\alpha(p)&\equiv  \langle \hat b_{\alpha}(-p)\, \hat d(p)\rangle\\
	&= ic_{48}\,p_\alpha\nonumber\\
	\Delta^{_{(13)}}_{\alpha,\rho}(p)&\equiv  \langle \hat d_{\alpha}(-p)\, \hat d_{\rho}(p)\rangle\\
	&=c_{49}\,\eta_{\alpha\rho}+c_{50}\,p_\alpha p_\rho+ic_{51}\,p^\lambda\epsilon_{\alpha\lambda\rho}\nonumber\\
	\Delta^{_{(14)}}_\alpha(p)&\equiv \langle \hat d_{\alpha}(-p)\, \hat d(p)\rangle\\
	&= ic_{52}\,p_\alpha\nonumber\\
	\Delta^{_{(15)}}(p)&\equiv \langle \hat d(-p)\, \hat d(p)\rangle\ ,
	\end{align}
expanded on the following basis of tensors
	\bea
	A^{(0)}_{\alpha\beta,\rho\sigma}  &=& \frac{1}{2}(\eta_{\alpha\rho}\eta_{\beta\sigma}+\eta_{\alpha\sigma}\eta_{\beta\rho}) 
	\label{E-A0}\\
	A^{(1)}_{\alpha\beta,\rho\sigma} &=& 
	\eta_{\alpha\rho}p_\beta p_\sigma +  
	\eta_{\alpha\sigma}p_\beta p_\rho +
	\eta_{\beta\rho}p_\alpha p_\sigma +
	\eta_{\beta\sigma}p_\alpha p_\rho
	\label{E-A1}\\
	A^{(2)}_{\alpha\beta,\rho\sigma} &=& 
	\eta_{\alpha\beta} p_\rho p_\sigma +
	\eta_{\rho\sigma} p_\alpha p_\beta
	\label{E-A2} \\
	A'^{(2)}_{\alpha\beta,\rho\sigma} &=& 
	\eta_{\alpha\beta} p_\rho p_\sigma 
	\label{E-A2'} \\
	B^{(2)}_{\alpha\beta,\rho\sigma} &=& 
	p_\alpha p_\beta\eta_{\rho\sigma} 
	\label{E-B2} \\
	A^{(3)}_{\alpha\beta,\rho\sigma} &=& 
	\eta_{\alpha\beta}\eta_{\rho\sigma}
	\label{E-A3}\\
	A^{(4)}_{\alpha\beta,\rho\sigma} &=& 
	p_\alpha p_\beta p_\rho p_\sigma
	\label{E-A4}\\
	A^{(5)}_{\alpha\beta,\rho\sigma} &=& 
	ip^\lambda (
	\epsilon_{\alpha\lambda\rho} \eta_{\sigma\beta} +
	\epsilon_{\beta\lambda\rho} \eta_{\sigma\alpha} +
	\epsilon_{\alpha\lambda\sigma} \eta_{\rho\beta} +
	\epsilon_{\beta\lambda\sigma} \eta_{\rho\alpha}
	)
	\label{E-A5}\\
	A^{(6)}_{\alpha\beta,\rho\sigma} &=&
	 ip^\lambda (
	\epsilon_{\alpha\lambda\rho} p_\sigma p_\beta +
	\epsilon_{\alpha\lambda\sigma} p_\rho p_\beta +
	\epsilon_{\beta\lambda\rho} p_\sigma p_\alpha +
	\epsilon_{\beta\lambda\sigma} p_\rho p_\alpha
	)\ ,
	\label{E-A6}
	\eea
where, since
	\be
	\Delta_{AP}=\Delta^\dag_{PA}\ ,
	\ee
the symmetries
	\be
	Z^{_{(i)}}_{M,P}(p)=Z^{_{(i)}*}_{P,M}(p)=Z^{_{(i)}*}_{M,P}(-p) \quad\mbox{for}\ Z^{_{(i)}}=\{\Delta^{_{i=1,6,10,11,13}}\ ,\ A^{_{i=0,...,6}}\}
	\ee
$i.e.$ when $M=\mu\nu$ and $P=\rho\sigma$ or $M=\mu$ and $P=\rho$, and
	\be
	Z^{_{(i)}}_{\alpha\beta,P}(p)=Z^{_{(i)}}_{\beta\alpha,P}(p)=Z^{_{(i)}*}_{\alpha\beta,P}(-p)\quad\mbox{for}\ Z^{_{(i)}}=\{\Delta^{_{i=1,...,9}}\ ,\ A'^{(2)},B^{(2)}\}
	\ee
$i.e.$ when $P=\{\rho\sigma\ ,\ \sigma\rho\ ,\ \rho\ ,\ \cdot\ \}$,  have been taken into account. From the invertibility condition \eqref{E-KD=I} we get the following system of equations
	\begin{align}
	\tfrac1 2 G^{\mu\nu,\alpha\beta}\Delta^{_{(2)}*}_{\rho\sigma,\alpha\beta}+\tfrac 1 2 G^{\mu\nu,\alpha}\Delta^{_{(3)}*}_{\rho\sigma,\alpha}+\tfrac 1 2 G^{\mu\nu}\Delta^{_{(5)}*}_{\rho\sigma}&=\mathcal{I}^{\mu\nu}_{\rho\sigma}(0)\label{E-eq1}\\
	\tfrac1 2 G^{\mu\nu,\alpha\beta}\Delta^{_{(6)}}_{\alpha\beta,\rho\sigma}+\tfrac 1 2 G^{\mu\nu,\alpha}\Delta^{_{(7)}*}_{\rho\sigma,\alpha}+\tfrac 1 2 G^{\mu\nu}\Delta^{_{(9)}*}_{\rho\sigma}&=0\label{E-eq2}\\
	\tfrac1 2 G^{\mu\nu,\alpha\beta}\Delta^{_{(7)}}_{\alpha\beta,\rho}+\tfrac 1 2 G^{\mu\nu,\alpha}\Delta^{_{(10)}}_{\alpha,\rho}+\tfrac 1 2 G^{\mu\nu}\Delta^{_{(12)}*}_\rho&=0\label{E-eq3}\\
	\tfrac1 2 G^{\mu\nu,\alpha\beta}\Delta^{_{(8)}}_{\alpha\beta,\rho}+\tfrac 1 2 G^{\mu\nu,\alpha}\Delta^{_{(11)}}_{\alpha,\rho}+\tfrac 1 2 G^{\mu\nu}\Delta^{_{(14)}*}_\rho&=0\label{E-eq4}\\
	\tfrac1 2 G^{\mu\nu,\alpha\beta}\Delta^{_{(9)}}_{\alpha\beta}+\tfrac 1 2 G^{\mu\nu,\alpha}\Delta^{_{(12)}}_\alpha+\tfrac 1 2 G^{\mu\nu}\Delta^{_{(15)}}&=0\label{E-eq5}\\[5px]
	\tfrac1 2 G^{\mu\nu,\alpha\beta}\Delta^{_{(1)}}_{\alpha\beta,\rho\sigma}+\tfrac 1 2 G^{*\mu\nu,\alpha}_{^{(\kappa_0,\kappa_1)}}\Delta^{_{(4)}*}_{\rho\sigma,\alpha}&=0\label{E-eq6}\\
	\tfrac1 2 G^{\mu\nu,\alpha\beta}\Delta^{_{(2)}}_{\alpha\beta,\rho\sigma}+\tfrac 1 2 G^{*\mu\nu,\alpha}_{^{(\kappa_0,\kappa_1)}}\Delta^{_{(8)}*}_{\rho\sigma,\alpha}&=\mathcal{I}^{\mu\nu}_{\rho\sigma}(\lambda)\label{E-eq7}\\
	\tfrac1 2 G^{\mu\nu,\alpha\beta}\Delta^{_{(3)}}_{\alpha\beta,\rho}+\tfrac 1 2 G^{*\mu\nu,\alpha}_{^{(\kappa_0,\kappa_1)}}\Delta^{_{(11)}*}_{\rho,\alpha}&=0\label{E-eq8}\\
	\tfrac1 2 G^{\mu\nu,\alpha\beta}\Delta^{_{(4)}}_{\alpha\beta,\rho}+\tfrac 1 2 G^{*\mu\nu,\alpha}_{^{(\kappa_0,\kappa_1)}}\Delta^{_{(13)}}_{\alpha,\rho}&=0\label{E-eq9}\\
	\tfrac1 2 G^{\mu\nu,\alpha\beta}\Delta^{_{(5)}}_{\alpha\beta}+\tfrac 1 2 G^{*\mu\nu,\alpha}_{^{(\kappa_0,\kappa_1)}}\Delta^{_{(14)}}_\alpha&=0\label{E-eq10}\\[5px]
	\tfrac 1 2 G^{*\alpha\beta,\mu}\Delta^{_{(1)}}_{\alpha\beta,\rho\sigma}+\tfrac 1 2 G^{\mu,\alpha}\Delta^{_{(4)}*}_{\rho\sigma,\alpha}&=0\label{E-eq11}\\
	\tfrac 1 2 G^{*\alpha\beta,\mu}\Delta^{_{(2)}}_{\alpha\beta,\rho\sigma}+\tfrac 1 2 G^{\mu,\alpha}\Delta^{_{(8)}*}_{\rho\sigma,\alpha}&=0\label{E-eq12}\\
	\tfrac 1 2 G^{*\alpha\beta,\mu}\Delta^{_{(3)}}_{\alpha\beta,\rho}+\tfrac 1 2 G^{\mu,\alpha}\Delta^{_{(11)}*}_{\rho,\alpha}&=\delta^\mu_\rho\label{E-eq13}\\
	\tfrac 1 2 G^{*\alpha\beta,\mu}\Delta^{_{(4)}}_{\alpha\beta,\rho}+\tfrac 1 2 G^{\mu,\alpha}\Delta^{_{(13)}}_{\alpha,\rho}&=0\label{E-eq14}\\
	\tfrac 1 2 G^{*\alpha\beta,\mu}\Delta^{_{(5)}}_{\alpha\beta}+\tfrac 1 2 G^{\mu,\alpha}\Delta^{_{(14)}}_\alpha&=0\label{E-eq15}\\[5px]
	\tfrac 1 2 G^{\alpha\beta,\mu}_{^{(\kappa_0,\kappa_1)}}\Delta^{_{(2)}*}_{\rho\sigma,\alpha\beta}+\tfrac 1 2 G^{*\alpha,\mu}\Delta^{_{(3)}*}_{\rho\sigma,\alpha}&=0\label{E-eq16}\\
	\tfrac 1 2 G^{\alpha\beta,\mu}_{^{(\kappa_0,\kappa_1)}}\Delta^{_{(6)}}_{\alpha\beta,\rho\sigma}+\tfrac 1 2 G^{*\alpha,\mu}\Delta^{_{(7)}*}_{\rho\sigma,\alpha}&=0\label{E-eq17}\\
	\tfrac 1 2 G^{\alpha\beta,\mu}_{^{(\kappa_0,\kappa_1)}}\Delta^{_{(7)}}_{\alpha\beta,\rho}+\tfrac 1 2 G^{*\alpha,\mu}\Delta^{_{(10)}}_{\alpha,\rho}&=0\label{E-eq18}\\
	\tfrac 1 2 G^{\alpha\beta,\mu}_{^{(\kappa_0,\kappa_1)}}\Delta^{_{(8)}}_{\alpha\beta,\rho}+\tfrac 1 2 G^{*\alpha,\mu}\Delta^{_{(11)}}_{\alpha,\rho}&=\delta^\mu_\rho\label{E-eq19}\\
	\tfrac 1 2 G^{\alpha\beta,\mu}_{^{(\kappa_0,\kappa_1)}}\Delta^{_{(9)}}_{\alpha\beta}+\tfrac 1 2 G^{*\alpha,\mu}\Delta^{_{(12)}}_\alpha&=0\label{E-eq20}\\[5px]
	\tfrac1 2 G^{\alpha\beta}\Delta^{_{(1)}}_{\alpha\beta,\rho\sigma}&=0\label{E-eq21}\\
	\tfrac1 2 G^{\alpha\beta}\Delta^{_{(2)}}_{\alpha\beta,\rho\sigma}&=0\label{E-eq22}\\
	\tfrac1 2 G^{\alpha\beta}\Delta^{_{(3)}}_{\alpha\beta,\rho}&=0\label{E-eq23}\\
	\tfrac1 2 G^{\alpha\beta}\Delta^{_{(4)}}_{\alpha\beta,\rho}&=0\label{E-eq24}\\
	\tfrac1 2 G^{\alpha\beta}\Delta^{_{(5)}}_{\alpha\beta}&=1\ .\label{E-eq25}
	\end{align}

Notice that saturating the $\rho\sigma$ indices in \eqref{E-eq7} and using the definitions \eqref{E-<ah>} and \eqref{E-<bh>} we have
	\be
		\begin{split}
		(1+3\lambda)\eta^{\mu\nu}&=\tfrac1 2 G^{\mu\nu,\alpha\beta}\eta^{\rho\sigma}\Delta^{_{(2)}}_{\alpha\beta,\rho\sigma}+\tfrac 1 2 G^{*\mu\nu,\alpha}_{^{(\kappa_0,\kappa_1)}}\eta^{\rho\sigma}\Delta^{_{(8)}*}_{\rho\sigma,\alpha}\\
		&=\tfrac1 2 G^{\mu\nu,\alpha\beta}\eta^{\rho\sigma}\langle\hat a_{\alpha\beta}(-p)\,\hat{b}_{\rho\sigma}(p)\rangle+\tfrac 1 2 G^{*\mu\nu,\alpha}_{^{(\kappa_0,\kappa_1)}}\eta^{\rho\sigma}\langle\hat d_\alpha(-p)\,\hat{b}_{\rho\sigma}(p)\rangle\\
		&=\tfrac1 2 G^{\mu\nu,\alpha\beta}\langle\hat a_{\alpha\beta}(-p)\,\hat{b}(p)\rangle+\tfrac 1 2 G^{*\mu\nu,\alpha}_{^{(\kappa_0,\kappa_1)}}\langle\hat d_\alpha(-p)\,\hat{b}(p)\rangle\ ,
		\end{split}
	\ee
which must vanish if the theory does not depend on the trace $\hat{b}(p)$, which would imply $\lambda=-\frac1 3$. This further justifies the introduction of the $\lambda$  parameter in the identity. From \eqref{E-eq1}-\eqref{E-eq25} the following equations are recovered through the multiplication rules of the basis \eqref{E-A0}-\eqref{E-A6}, which can be found in \cite{Bertolini:2024yur}
\begin{align}
2 p^2c_{13}+ 1 &= 0\qquad\mbox{from \eqref{E-eq1}}\\
3 c_{13}-p^2c_{14}+ c_{18}&= 0 \nonumber\\
-c_{13}- \tfrac1 2  p^2c_{16}- c_{18}- \tfrac1 2 k_1 p^2c_{24}&= 0 \nonumber\\
-c_{13}+ \tfrac1 2  c_{15}- \tfrac1 2 k_0 c_{23}&= 0 \nonumber\\
c_{13}p^2- \tfrac1 2  p^2c_{15}- \tfrac1 2 k_1 p^2c_{23} &= 0 \nonumber\\
c_{14}+ \tfrac1 2  c_{16}- \tfrac1 2 k_0 c_{24}&= 0 \nonumber\\
c_{7}&= 0 \nonumber\\
c_{7}- c_{17}&= 0 \nonumber\\
[10px]
c_{30}&= 0\qquad\mbox{from \eqref{E-eq2}}\\
3 c_{30}- p^2c_{31} + c_{35}&= 0 \nonumber\\
c_{30}+ \tfrac1 2  p^2c_{33}+ c_{35}+ \tfrac1 2 k_1 p^2c_{41}&= 0 \nonumber\\
-c_{30}+ \tfrac1 2  c_{32}- \tfrac1 2 k_0 c_{40}&= 0 \nonumber\\
c_{30}p^2- \tfrac1 2  p^2c_{32}- \tfrac1 2 k_1 p^2c_{40}&= 0 \nonumber\\
c_{31}+ \tfrac1 2  c_{33}- \tfrac1 2 k_0 c_{41}&= 0 \nonumber\\
c_{25}&= 0 \nonumber\\
c_{26}- c_{34}&= 0 \nonumber\\
[10px]
p^2c_{34}+ c_{42}&= 0 \qquad\mbox{from \eqref{E-eq3}}\\
c_{34}- c_{43}- k_0 c_{48}&= 0 \nonumber\\
c_{35}+ c_{44}&= 0 \nonumber\\
-c_{42}- p^2c_{43}+ k_1 p^2c_{48}&= 0 \nonumber\\
[10px]
p^2c_{38}+ c_{45}&= 0 \qquad\mbox{from \eqref{E-eq4}}\\
 c_{38}-c_{46}- k_0 c_{52}&= 0 \nonumber\\
c_{39}+ c_{47}&= 0 \nonumber\\
-c_{45}- p^2c_{46}+ k_1 p^2c_{52}&= 0 \nonumber\\
[10px]
c_{48}p^2- \Delta^{_{(15)}}  k_1 p^2&= 0 \qquad\mbox{from \eqref{E-eq5}}\\
c_{48}+ \Delta^{_{(15)}}  k_0 &= 0 \nonumber\\
[10px]
c_{5}&= 0 \qquad\mbox{from \eqref{E-eq6}}\\
-3 c_{5}+ p^2c_{6}+ \kappa_0 c_{22}&= 0 \nonumber\\
2 c_{5}+ \kappa_1 (p^2c_{19}+ 2 c_{22}) &= 0 \nonumber\\
2 c_{5}+ \kappa_0 c_{19}&= 0 \nonumber\\
-2 c_{5}+ \kappa_1 p^2c_{19}&= 0 \nonumber\\
-2 c_{6}+ \kappa_0 c_{20}&= 0 \nonumber\\
c_{0}&= 0 \nonumber\\
c_{1}+ \kappa_0 c_{21}&= 0 \nonumber\\
[10px]
2 c_{13}p^2+1 &= 0\qquad\mbox{from \eqref{E-eq7}}\\
-3 c_{13}+ p^2c_{14}+ \kappa_0 c_{39}&= 0 \nonumber\\
c_{13}+ \tfrac1 2 \kappa_1 p^2c_{37}+ \kappa_1 c_{39}&= 0 \nonumber\\
c_{13}+ \tfrac1 2 \kappa_0 c_{36}&= 0 \nonumber\\
p^2c_{13}- \tfrac1 2 \kappa_1 p^2c_{36}-\lambda &= 0 \nonumber\\
-c_{14}+ \tfrac1 2 \kappa_0 c_{37}&= 0 \nonumber\\
c_{7}&= 0 \nonumber\\
c_{8}+ \kappa_0 c_{38}&= 0 \nonumber\\
[10px]
-p^2c_{17}+ \kappa_0 c_{45}&= 0\qquad\mbox{from \eqref{E-eq8}}\\
c_{17}+ \kappa_0 c_{46}&= 0 \nonumber\\
c_{18}- \kappa_0 c_{47}&= 0 \nonumber\\
\kappa_1 (c_{45}+ p^2c_{46}) &= 0 \nonumber\\
[10px]
-p^2c_{21}+ \kappa_0 c_{49}&= 0 \qquad\mbox{from \eqref{E-eq9}}\\
c_{21}+ \kappa_0 c_{50}&= 0 \nonumber\\
c_{22}- \kappa_0 c_{51}&= 0 \nonumber\\
\kappa_1 (c_{49}+ p^2c_{50}) &= 0 \nonumber\\
[10px]
\kappa_0 c_{52}&= 0 \qquad\mbox{from \eqref{E-eq10}}\\
\kappa_1 c_{52}&= 0 \nonumber\\
[10px]
c_{0}+ 2 c_{1}p^2+2\kappa_2 p^2c_{21}&= 0 \qquad\mbox{from \eqref{E-eq11}}\\
-c_{1}- c_{2}-\kappa_2 c_{21}&= 0 \nonumber\\
c_{3}- c_{0}&= 0 \nonumber\\
c_{5}+ p^2c_{6}+\kappa_2 c_{22}&= 0 \nonumber\\
[10px]
c_{7}+ 2 c_{8}p^2+2\kappa_2 p^2c_{38}&= 0 \qquad\mbox{from \eqref{E-eq12}}\\
-c_{8}- c_{9}-\kappa_2 c_{38}&= 0 \nonumber\\
c_{11}- c_{7}&= 0 \nonumber\\
c_{13}+ p^2c_{14}+\kappa_2 c_{39}&= 0 \nonumber\\
[10px]
2 - p^2c_{18}+\kappa_2 p^2c_{47}&= 0 \qquad\mbox{from \eqref{E-eq13}}\\
-2c_{15}- c_{18}+\kappa_2 c_{47}&= 0 \nonumber\\
p^2c_{17}-\kappa_2c_{45}&=0\nonumber\\
[10px]
p^2c_{22}-\kappa_2 p^2c_{51}&= 0 \qquad\mbox{from \eqref{E-eq14}}\\
-2c_{19}- c_{22}+\kappa_2 c_{51}&= 0 \nonumber\\
c_{20}-\kappa_2c_{49}&=0\nonumber\\
[10px]
c_{23}&= 0 \qquad\mbox{from \eqref{E-eq15}}\\
[10px]
\tfrac1 2 \kappa_0 c_{7}+ \kappa_0 p^2c_{8}-\kappa_2 p^2c_{17}&= 0 \qquad\mbox{from \eqref{E-eq16}}\\
\kappa_0 c_{13}+ \kappa_0 p^2c_{14}-\kappa_2  c_{18}&= 0 \nonumber\\
2 (\kappa_0 + 2 \kappa_1) c_{8}+ (\kappa_0 + 3 \kappa_1) c_{10}+ (\kappa_0 + \kappa_1) p^2c_{12}+2\kappa_2 c_{17}&= 0 \nonumber\\
(\kappa_0 + \kappa_1) p^2c_{9}+ (\kappa_0 + 3 \kappa_1) c_{11}+ \kappa_1 c_{7}&= 0 \nonumber\\
[10px]
\tfrac1 2 \kappa_0 c_{25}+ \kappa_0 p^2c_{26}-\kappa_2 p^2c_{34}&= 0 \qquad\mbox{from \eqref{E-eq17}}\\
\kappa_0 c_{30}+ \kappa_0 p^2c_{31} -\kappa_2  c_{35}&= 0 \nonumber\\
2 (\kappa_0 + 2 \kappa_1) c_{26}+ (\kappa_0 + 3 \kappa_1) c_{27}+ (\kappa_0 + \kappa_1) p^2c_{29}+2\kappa_2 c_{34}&= 0 \nonumber\\
(\kappa_0 + \kappa_1) p^2c_{27}+ (\kappa_0 + 3 \kappa_1) c_{28}+ \kappa_1 c_{25}&= 0 \nonumber\\
[10px]
(\kappa_0 + 3 \kappa_1) c_{32}+ (\kappa_0 + \kappa_1) p^2c_{33}+ (\kappa_0 + 2 \kappa_1) c_{35}-\kappa_2 c_{44}&= 0 \qquad\mbox{from \eqref{E-eq18}}\\
\kappa_0 p^2c_{34}+\kappa_2 c_{42}&= 0 \nonumber\\
\kappa_0 c_{35}+\kappa_2 c_{44}&= 0 \nonumber\\
[10px]
(\kappa_0 + 3 \kappa_1) c_{36}+ (\kappa_0 + \kappa_1) p^2c_{37}+ (\kappa_0 + 2 \kappa_1) c_{39}-\kappa_2 c_{47}&= 0 \qquad\mbox{from \eqref{E-eq19}}\\
\kappa_0 p^2c_{38}+\kappa_2 c_{45}&= 0 \nonumber\\
\kappa_0 p^2c_{39}+\kappa_2 p^2c_{47}+2 &= 0 \nonumber\\
[10px]
(\kappa_0 + 3 \kappa_1) c_{40}+ (\kappa_0 + \kappa_1) p^2c_{41}&= 0 \qquad\mbox{from \eqref{E-eq20}}\\
[10px]
k_0 c_{0}+ 4 (k_0 + k_1) p^2c_{1}+ (k_0 + 3 k_1) p^2c_{2}+ (k_1 +  k_0) p^4 c_{4}&= 0\qquad\mbox{from \eqref{E-eq21}}\\
k_1 p^2c_{0}+ (k_0 + k_1) p^4 c_{2}+ (k_0 + 3 k_1) p^2c_{3}&= 0 \nonumber\\
[10px]
k_0 c_{7}+  4 (k_0 + k_1) p^2c_{8}+ (k_0 + 3 k_1) p^2c_{9}+ (k_1 +   k_0) p^4 c_{12}&= 0\qquad\mbox{from \eqref{E-eq22}}\\
k_1 p^2c_{7}+ (k_0 + k_1) p^4 c_{10}+ (k_0 + 3 k_1) p^2c_{11}&= 0 \nonumber\\
[10px]
(k_0 + 3 k_1) c_{15}+ (k_0 + k_1) p^2c_{16}+ 2 (k_0 + k_1) c_{18}&= 0 \qquad\mbox{from \eqref{E-eq23}}\\
[10px]
(k_0 + 3 k_1) c_{19}+ (k_0 + k_1) p^2c_{20}+ 2 (k_0 + k_1) c_{22}&= 0 \qquad\mbox{from \eqref{E-eq24}}\\
[10px]
(k_0 + 3 k_1) p^2c_{23}+ (k_0 + k_1) p^4 c_{24}+ 2 &= 0\qquad\mbox{from \eqref{E-eq25}}
\end{align}
\allowdisplaybreaks[0]
 Solutions to this system of equations are
	\begin{align}
	\lambda&=-\frac{1}{3}&\kappa_1&=-\frac{1}{3}\kappa_0&\\
	c_{13}&=-\frac{1}{2p^2} &c_{14}&=\frac{\kappa_0+3\kappa_2}{\kappa_0-\kappa_2}\frac{1}{2p^4}&c_{15}&=-\frac{1}{p^2}\\
	c_{16}&=-\left(\frac{k_0-k_1}{k_0+k_1}+2\frac{\kappa_0+\kappa_2}{\kappa_0-\kappa_2}\right)
\frac{1}{p^4}&c_{18}&=\frac{2\kappa_0}{\kappa_0-\kappa_2}\frac{1}{p^2}&	c_{24}&=-\frac{2}{k_0+k_1}\frac{1}{p^4} \\
	c_{36}&=\frac{1}{\kappa_0p^2}&c_{37}&=\frac{\kappa_0+3\kappa_2}{\kappa_0(\kappa_0-\kappa_2)}\frac{1}{p^4}&c_{39}&=-\frac{2}{\kappa_0-\kappa_2}\frac{1}{p^2}\\	
	c_{47}&=\frac{2}{\kappa_0-\kappa_2}\frac{1}{p^2}\ ,
	\end{align}
and	
	\be
	c_i=\Delta^{_{(15)}}=0\quad\mbox{for}\ i=\{\mbox{0-12,17,19-23,25-35,38,40-46,48-52}\}\ ,\\
	\ee
which signal poles at
	\be
	\kappa_0=0\quad;\quad\kappa_0=\kappa_2\quad;\quad k_1=-k_0\ .
	\ee
Non-trivial propagators are thus the following
	\begin{align}
	\Delta^{_{(2)}}_{\alpha\beta,\rho\sigma}(p)&\equiv \langle \hat a_{\alpha\beta}(-p)\, \hat{b}_{\rho\sigma}(p)\rangle\\
	&=\frac{1}{2p^2}\left(-A^{(5)}_{\alpha\beta,\rho\sigma}+\frac{\kappa_0+3\kappa_2}{\kappa_0-\kappa_2}\frac{1}{p^2}A^{(6)}_{\alpha\beta,\rho\sigma}\right)\nonumber\\
&=	-\frac{i}{2p^2}p^\lambda\bigg[\left(	\epsilon_{\alpha\lambda\rho}  t_{\sigma\beta} +	\epsilon_{\beta\lambda\rho}  t_{\sigma\alpha} +	\epsilon_{\alpha\lambda\sigma}  t_{\rho\beta} +	\epsilon_{\beta\lambda\sigma}  t_{\rho\alpha}\right)+\nonumber\\
	&\quad\left.-\frac{4\kappa_2}{\kappa_0-\kappa_2}	  \left(\epsilon_{\alpha\lambda\rho}\tfrac{p_\sigma p_\beta}{p^2}  +\epsilon_{\alpha\lambda\sigma} \tfrac{p_\rho p_\beta}{p^2} +	\epsilon_{\beta\lambda\rho}\tfrac{ p_\sigma p_\alpha}{p^2} +\epsilon_{\beta\lambda\sigma} \tfrac{p_\rho p_\alpha}{p^2}\right)\right]\nonumber\\[5px]
	\Delta^{_{(3)}}_{\alpha\beta,\rho}(p)&\equiv \langle \hat a_{\alpha\beta}(-p)\, \hat b_{\rho}(p)\rangle\\
	&=-\frac{i}{p^2}\left[\eta_{\alpha\beta}p_\rho+i\left(\frac{k_0-k_1}{k_0+k_1}+2\frac{\kappa_0+\kappa_2}{\kappa_0-\kappa_2}\right)\frac{p_\alpha p_\beta}{p^2} p_\rho-\frac{2\kappa_0}{\kappa_0-\kappa_2}\left(\eta_{\alpha\rho}p_\beta+\eta_{\beta\rho}p_\alpha\right)\right]\nonumber\\
	&= \frac{i}{p^2}\left[\frac{2\kappa_0}{\kappa_0-\kappa_2}\left(t_{\alpha\rho}p_\beta+t_{\beta\rho}p_\alpha\right)-\frac{1}{k_0+k_1}\left(k_0\, t_{\alpha\beta}p_\rho+k_1\, \tilde t_{\alpha\beta}p_\rho\right)\right]\nonumber\\[5px]
	\Delta^{_{(5)}}_{\alpha\beta}(p)&\equiv \langle \hat a_{\alpha\beta}(-p)\, \hat d(p)\rangle= -\frac{2}{k_0+k_1}\frac{p_\alpha p_\beta}{p^4}\\[5px]
	\Delta^{_{(8)}}_{\alpha\beta,\rho}(p)&\equiv \langle \hat{b}_{\alpha\beta}(-p)\, \hat d_{\rho}(p)\rangle\\
	&=\frac{i}{p^2}\left[\frac{1}{\kappa_0}\eta_{\alpha\beta}p_\rho+\frac{\kappa_0+3\kappa_2}{\kappa_0(\kappa_0-\kappa_2)}\frac{p_\alpha p_\beta}{p^2} p_\rho-\frac{2}{\kappa_0-\kappa_2}\left(\eta_{\alpha\rho}p_\beta+\eta_{\beta\rho}p_\alpha\right)\right]\nonumber\\
	&= \frac{i}{p^2}\left[\frac{1}{\kappa_0}\tilde t_{\alpha\beta}p_\rho-\frac{2}{\kappa_0-\kappa_2}\left(t_{\alpha\rho}p_\beta+t_{\beta\rho}p_\alpha\right)\right]\nonumber\\[5px]
	\Delta^{_{(11)}}_{\alpha,\rho}(p)&\equiv  \langle \hat b_{\alpha}(-p)\, \hat d_{\rho}(p)\rangle=\frac{2i}{\kappa_0-\kappa_2}\frac{\epsilon_{\alpha\lambda\rho}p^\lambda}{p^2}\ ,
	\end{align}
where
	\be
	t_{\alpha\beta}\equiv \eta_{\alpha\beta}-\tfrac{p_\alpha p_\beta}{p^2}\quad;\quad\tilde t_{\alpha\beta}\equiv \eta_{\alpha\beta}-3\tfrac{p_\alpha p_\beta}{p^2}\ ,
	\ee
such that
	\be
	p^\alpha t_{\alpha\beta}=0\quad;\quad\eta^{\alpha\beta}\tilde t_{\alpha\beta}=0\,.
	\ee
It is thus immediate to observe that
	\be
	\eta^{\rho\sigma}\Delta^{_{(2)}}_{\alpha\beta,\rho\sigma}(p)=\langle \hat a_{\alpha\beta}(-p)\, \hat{b}(p)\rangle=0=\eta^{\alpha\beta}\Delta^{_{(8)}}_{\alpha\beta,\rho}(p)=\langle \hat{b}(-p)\, \hat d_{\rho}(p)\rangle\ ,
	\ee
which confirms that the trace $b(x)$ has no role in the theory.

\newpage


\end{document}